Kepler-22b: A 2.4 Earth-radius Planet in the Habitable Zone of a Sun-like Star†

William J. Borucki[0,1], David G. Koch[1], Natalie Batalha[3], Stephen T. Bryson[1], Douglas A. Caldwell[6], Jørgen Christensen-Dalsgaard[7,20], William D. Cochran[8], Edna DeVore[6], Thomas N. Gautier III[11], John C. Geary[10], Ronald Gilliland[12], Alan Gould[13], Steve B. Howell[14], Jon M. Jenkins[6], David W. Latham[10], Jack J. Lissauer[1], Geoffrey W. Marcy[2], Jason Rowe[1], Dimitar Sasselov[10], Alan Boss[4], David Charbonneau[10], David Ciardi[22], Guillermo Torres[10], Francois Fressin[10], Lisa Kaltenegger[30], Laurance Doyle[6], Andrea K. Dupree[10], Eric B. Ford[16], Jonathan Fortney[17], Matthew J. Holman[10], Jason A. Steffen[9], Fergal Mullally[6], Martin Still[27], Jill Tarter[6], Sarah Ballard[10], Lars A. Buchhave[10], Josh Carter[18], Jessie L. Christiansen[6], Brice-Olivier Demory[18], Jean-Michel Désert[10], Courtney Dressing[10], Michael Endl[8], Daniel Fabrycky[17], Debra Fischer[19], Michael R. Haas[1], Christopher Henze[1], Elliott Horch[24], Andrew W. Howard[2], Howard Isaacson[2], Hans Kjeldsen[7], John Asher Johnson[11], Todd Klaus[21], Jeffery Kolodziejczak[25], Thomas Barclay[27], Jie Li[6], Søren Meibom[10,], Andrej Prsa[26], Samuel N. Quinn[10], Elisa V. Quintana[6], Paul Robertson[8], William Sherry[14], Avi Shporer[5], Peter Tenenbaum[6], Susan E. Thompson[6], Joseph D. Twicken[6], Jeffrey Van Cleve[6], William F. Welsh[20], Sarbani Basu[19], Bill Chaplin[29], Andrea Miglio[29], Steve Kawaler[31], Torben Arentoft[7], Dennis Stello[32], Travis S. Metcalfe[33], GrahamVerner[29], Christoffer Karoff[7], Mia Lundkvist[7], Mikkel Lund[7], Rasmus Handberg[7], Yvonne Elsworth[29], Saskia Hekker[34,29], Daniel Huber[32], Timothy R. Bedding[32]

[0]Correspondence should be addressed to: William Borucki, William.J.Borucki@nasa.gov
[1]NASA Ames Research Center, Moffett Field, CA 94035, USA
[2]University of California, Berkeley, CA, 94720, USA
[3]San Jose State University, San Jose, CA, 95192, USA
[4]Carnegie Institution of Washington, Washington, DC 20015 USA
[6]SETI Institute, Mountain View, CA, 94043, USA
[7]Aarhus University, Aarhus, Denmark
[8]McDonald Observatory, University of Texas at Austin, Austin, TX, 78712, USA
[9]Fermilab, Batavia, IL 60510, USA
[10]Harvard-Smithsonian Center for Astrophysics, Cambridge, MA, 02138, USA
[11]Jet Propulsion Laboratory, Calif. Institute of Technology, Pasadena, CA, 91109, USA
[12]Space Telescope Science Institute, Baltimore, MD, 21218, USA
[13] Lawrence Hall of Science, Berkeley, CA 94720, USA
[14]NOAO, Tucson, AZ 85719 USA
[15]University of Arizona, Steward Observatory, Tucson, AZ 85721, USA
[16]Univ. of Florida, Gainesville, FL, 32611 USA
[17]Univ. of Calif., Santa Cruz, CA 95064 USA
[18]MIT, Cambridge, MA 02139 USA
[19]Yale University, New Haven, CT 06520 USA
[20]High Altitude Observatory, NCAR, Boulder, CO, 80307, USA
[21]Orbital Sciences Corp., Mountain View, CA 94043 USA
[22]Exoplanet Science Institute/Caltech, Pasadena, CA 91125 USA
[23]University Affiliated Research Center, University of California, Santa Cruz, CA 95064 USA
[24]Southern Connecticut State University, New Haven, CT 06515 USA
[25]MSFC, Huntsville, AL 35805 USA
[26]Villanova University, Villanova, PA 19085 USA
[27]Bay Area Environmental Research Institute/ Moffett Field, CA 94035, USA
[28]University of Hertfordshire, Hatfield, UK
[29]University of Birmingham, Birmingham, UK
[30]Max-Planck-Institut fur Astronomie, Heidelberg, D 69117, Germany
[31]Iowa State University, Ames, IA 50011, USA
[32]University of Sydney, NSW, Australia

[33]White Dwarf Research Corporation, Boulder, CO 80301, USA
[34]University of Amsterdam, Amsterdam, The Netherlands

## ABSTRACT

A search of the time-series photometry from NASA's Kepler spacecraft reveals a transiting planet candidate orbiting the 11th magnitude G5 dwarf KIC 10593626 with a period of 290 days. The characteristics of the host star are well constrained by high-resolution spectroscopy combined with an asteroseismic analysis of the Kepler photometry, leading to an estimated mass and radius of 0.970 +/- 0.060 $M_\odot$ and 0.979 +/- 0.020 $R_\odot$. The depth of 492 ± 10ppm for the three observed transits yields a radius of 2.38 +/- 0.13 Re for the planet. The system passes a battery of tests for false positives, including reconnaissance spectroscopy, high-resolution imaging, and centroid motion. A full BLENDER analysis provides further validation of the planet interpretation by showing that contamination of the target by an eclipsing system would rarely mimic the observed shape of the transits. The final validation of the planet is provided by 16 radial velocities obtained with HIRES on Keck 1 over a one year span. Although the velocities do not lead to a reliable orbit and mass determination, they are able to constrain the mass to a 3σ upper limit of 124 $M_\oplus$, safely in the regime of planetary masses, thus earning the designation Kepler-22b. The radiative equilibrium temperature is 262K for a planet in Kepler-22b's orbit. Although there is no evidence that Kepler-22b is a rocky planet, it is the first confirmed planet with a measured radius to orbit in the Habitable Zone of any star other than the Sun.

## 1. INTRODUCTION

*Kepler* is a Discovery-class mission designed to determine the frequency of Earth-radius planets in and near the habitable zone (HZ) of solar-type stars (Borucki et al 2009, 2010a, Caldwell et al. 2010, Koch et al 2010a). Since its launch in 2009, over 1200 candidate planets have been discovered with sizes ranging from less than Earth to twice as large as Jupiter and with orbital periods from less than a day to more than a year. Confirming and validating these candidates as planets requires a lengthy process to avoid false positive events that would lead to inaccurate statistics of characteristics of the exoplanet population. Because of the large-amplitude radial velocity (RV) signatures, the first confirmations of *Kepler* planets were Jupiter mass objects in short period orbits (Borucki et al. 2010b**,** Dunham et al. 2010, Koch et al. 2010b, Latham et al. 2010). As the Mission duration has increased and the data analysis pipeline capability has improved, smaller candidates, candidates in longer period orbits, multi-planet systems, and circumbinary planets have been found (Batalha et al, 2011, Holman et al, 2010, Lissauer et al 2011a, Doyle et al. 2011). In this paper we describe the validation of the first *Kepler* planet found in the HZ of its host star.

The instrument is a wide field-of-view (115 square degrees) photometer comprised of a 0.95-meter effective aperture Schmidt telescope feeding an array of 42 CCDs which continuously and simultaneously monitors the brightness of up to 170,000 stars. A comprehensive discussion of the characteristics and on-orbit performance of the instrument and spacecraft is presented in Argabright et al. (2008), Van Cleve and Caldwell (2009), and Koch et al. (2010a). The statistical properties of stars targeted by *Kepler* are described by Batalha et al. (2010a). Data for up to 170, 000 stars are observed in ~30 minute integrations (long-cadence or LC) while data for up to 512 stars are also observed in ~ 1-minute integrations (58.85 second) termed short-cadence (SC). The SC photometry is described in Jenkins et al. (2010a) and Gilliland et al. (2010). The LC photometry of Kepler-22b used in the analyses reported here was acquired between 13 May 2009 and 14 March 2011 – Quarter 1 through Quarter 8.

Here, we report on the discovery of a 2.38-Earth-radius planet (Kepler 22b) orbiting the G-type main sequence star listed in the Kepler Input Catalog (KIC) (Brown et al. 2011) as KIC 10593626 (Kepler-22) at RA = $19^{hrs}$ $16^{m}$ $52.2^{s}$ and Dec = +47° 53' 4.2". At *Kepler* magnitude (Kp) = 11.664, the star is bright enough for asteroseismic analysis of its fundamental stellar properties using the high precision *Kepler* photometry. Kepler-22b was previously listed as KOI-87.01 in the list of *Kepler* candidates (Borucki et al. 2011). All data used for the analysis of Kepler-22b are publically available at the Multiple Mission Archive at Space Telescope Science Institute (MAST[1]).

The Kepler-22b data acquisition, photometry and transit detection are described in Section 2. The statistical tests performed to rule out false positives are described in Section 3. The subsequent ground-based observations, including high-precision Doppler measurements, leading to the validation[2] of Kepler-22b are described in Section 4. The determination of the values for the stellar parameters is discussed in Section 5.

Section 6 presents the false-positive scenarios that were investigated using BLENDER analysis (Fressin et al. 2011).

Section 7 describes the joint modeling of the light curve and RV observations to provide best estimates of the planet and stellar characteristics. Section 8 discusses the use of transit timing variations to search for other, non-transiting planets. Section 9 discusses the implications of Kepler-22b being inside the habitable zone (HZ) of its host star. Section 10 presents the summary.

## 2. KEPLER PHOTOMETRY

[1] http://archive.stsci.edu/kepler

[2] In the context of this paper we reserve the term "confirmation" for the unambiguous detection of the gravitational influence of the planet on its host star (e.g., the Doppler signal) to establish the planetary nature of the candidate; when this is not possible, as in the present case, we speak of "validation", which involves an estimate of the false alarm probability.

*2.1. Data Acquisition*

Over 31,000 LC observations were obtained between 13 May 2009 and 14 March 2011. SC data were also collected between 20 August 2009 and 14 March 2011). The SC data were essential for determining the fundamental stellar (and thus the planet) parameters from an asteroseismic analysis (p-mode detection) described in Section 4.9. Approximately 790,000 SC observations were collected in this time period. Both LC and SC data are used in our light curve analysis.

The largest systematic errors are due to long-term image motion (differential velocity aberration) and the thermal transients after safe mode events. After masking transit events, the measured relative standard deviation of the PDC[3]-corrected, long-cadence light curve is 62 ppm per LC cadence. The propagated formal uncertainty from the instrument and photon shot noise is computed for each flux measurement in the time-series. The mean of the LC noise estimates reported by the pipeline is 36 ppm. The 62 ppm total measurement noise includes instrument noise, shot noise, noise introduced by the data reduction, and stellar variability. Both simple-aperture photometry (SAP) and corrected simple-aperture photometry (PDCSAP) are available at the MAST.

*2.2. Light Curves*

The two upper panels in Figure 1 show the full 22-month time series before and after analysis and detrending by the *Kepler* pipeline, respectively (Twicken et al. 2010, Jenkins et al.

[3] Pre-search Data Conditioning algorithm

2010b). Intra-quarter fluxes were normalized by their median flux in order to reduce the magnitude of the flux discontinuities between quarters. The red curve in the bottom panel is the model fit to the data. For values of the transit and orbital parameters, see Table 1.

## 3. STATISTICAL TESTS TO RULE OUT FALSE POSITIVES

Astrophysical signals mimicking planet transits are routinely picked up by the *Kepler* data analysis pipeline. The large majority of such false positives can be identified via statistical tests performed on the *Kepler* data themselves– tests that are collectively referred to as *Data Validation*. Here we describe the Data Validation metrics – statistics which, taken alone, support the planet interpretation for Kepler-22b.

### *3.1. Binarity Tests*

The depth of each transit is checked for consistency with the global model; i.e., that there's no significant evidence for the presence of a doubled-period eclipsing binary. A statistically significant difference in the transit depths would be an indication of a diluted or grazing eclipsing binary system Batalha et al. (2010b). The transit events detected in the light curve of Kepler-22b are shown in the fourth panel of Figure 1 where it can be seen that all three show differences $< 1\sigma$ where $\sigma$ refers to the uncertainty in the fitted transit depths of 10 ppm. Although only 3 transits are available for this test, there is no evidence of a secondary eclipse and the result of the binarity test is consistent with the planet interpretation.

*3.2. Photocenter Tests*

We check to see whether the transit signal is due to a source other than Kepler-22b using two methods. The first method measures the center-of-light distribution in the photometric aperture and will be referred to as flux-weighted centroids. The flux weighted centroid method measures the flux-weighted centroid of every observational cadence and fits the computed transit model multiplied by a constant amplitude to the observed flux-weighted centroid motion. The value of the constant that provides the best $\chi^2$ fit is taken to be the amplitude of the centroid motion. This constant is scaled by the transit depth to estimate the location of the transit source (Jenkins, et al. 2010c). The analysis shows no significant motion with a 1σ upper limit of 0.3″ (right-hand column ofTable 2).

The second technique uses the difference-image technique (Torres et al. 2011) and is referred to as pixel response function (PRF) fitting. The PRF fitting method fits the measured *Kepler* PRF (Bryson et al. 2010) to a difference image. This image is formed from the average in-transit and average out-of-transit (but near-transit) pixel images. The PRF fitted difference image centroid provides a direct measurement of the location of the transit signal in pixel space. This difference image centroid position is compared with the position of the PRF-fit centroid of the average out-of-transit image. Figure 2 shows an example of both techniques for Kepler-22b in Quarter 1, where the left column shows the observed data and difference image and the right column shows the reconstructed pixels based on the fitted PRF. The agreement between the columns shows that the fit was successful.

Both centroiding methods begin in pixel coordinates. To perform multi-quarter analysis, the pixel-level results are projected onto the sky in RA and Dec coordinates. In the case of flux-weighted centroids, this projection takes place during the $\chi^2$ fit. The PRF fitted centroids are computed quarter-by-quarter and the final results are also transformed to celestial coordinates. The quarterly PRF fitted results are then averaged (minimizing a robust $\chi^2$ fit to a constant position) to account for quarterly bias due to PRF error and possible crowding (crowding is a minor issue in the case of Kepler-22b, see Section 4.1). The final multi-quarter results for the PRF-fitted results are presented in Table 3 and shown in Figure 3.

Both centroid methods indicate that the location of the Kepler-22 transit source is less than 1.5σ from Kepler-22, a very strong indication that the transiting object is Kepler-22b. Both methods rule out a source greater than 0.9 arcsec away with a 3σ confidence level.

## 4. FOLLOW-UP OBSERVATIONS

Each of the three transit events identified in the light curve of Kepler-22b passes all of the Data Validation tests that might indicate the possibility of a false-positive as described in Section 3. To continue the validation process, a series of ground-based observations were initiated that included seeing-limited observations, active optics(AO) and speckle imaging to identify nearby stars in the photometric aperture, and a spectroscopic search for double-lined binary and background stars. Reconnaissance spectroscopy was employed to improve the accuracy of the stellar parameters in the KIC (Brown et al 2011).

The final steps included high SNR echelle spectroscopy with and without an iodine cell to compute stellar parameters, probe magnetic activity, measure line bisectors, and make high-precision Doppler measurements to obtain an upper limit to the mass of the planet.

### *4.1. Seeing-limited imaging*

Figure 4 shows a 1.2'x1.2' view centered on the Kepler-22 taken with the Lick Observatory 1-m Nickel Telescope to map nearby stars. The seeing is ~ 1.5". A companion is seen 5" to the South and is ~ 5 magnitudes fainter than the primary. An analysis of the nearby stars shows that they contribute contaminating fluxes ranging from 0.9% to 1.4%, depending on quarter. The flux light curves have been normalized to account for this contamination prior to planet search and characterization in the SOC pipeline.

### *4.2. AO Imaging*

Near-infrared adaptive optics imaging of Kepler-22b was obtained on the night of 03 July 2010 UT with the Palomar Hale 5m telescope and the Palomar High Angular Resolution Observer (PHARO) near-infrared camera (Hayward et al. 2001) behind the Palomar adaptive optics system (Troy et al. 2000). PHARO, a 1024×1024 HgCdTe infrared array, was utilized in 25.1 mas/pixel mode yielding a field of view of 25". Observations were performed in both the J ($\lambda$ = 1.25 μm) and Ks ($\lambda$ = 2.145 μm) filters. The data were collected in a standard 5-point quincunx dither pattern of 5" steps interlaced with an off-source (60" East) sky dither pattern. The integration time per source was 7.1 sec at Ks and 9.9 seconds at J. A total of 25 frames each were acquired at Ks and J for a total on-source integration time of 3 and 4 minutes, respectively. The individual frames were reduced with a custom set of IDL routines written for the Palomar High Angular Resolution Observer (PHARO) camera and were combined into a single final image. The adaptive optics system guided on the primary target itself; the widths of the central cores of the resulting point spread functions were FWHM = 0.10" at Ks and FWHM = 0.11" at J. The final coadded images at J and Ks are shown in Figure 5.

Three sources were detected within 10" of the target. The closest line-of-sight companion is separated from Kepler-22b by 5.5" to the south and has magnitudes of J = 16.48±0.03 mag and Ks = 16.01±0.02 mag. A second source was detected 5.5" to the northeast; that source has infrared magnitudes of J = 17.86±0.05 mag and Ks = 17.05±0.04 mag. Finally, a third source was marginally detected 9.5" to the southeast with infrared magnitudes of J = 20.2±0.5 mag and Ks = 18.9±0.4 mag. Based upon the Kp-Ks vs. J-Ks color-color relationships from Howell et al. (2011b), we derive Kp = 18.85 ± 0.1 mag, Kp = 19.9 ± 0.1 mag, and Kp = 21.8 ± 0.6 mag for 7, 8, and 10 magnitudes fainter than the primary target in the *Kepler* bandpass. Together, these stars dilute the light from Kepler-22 by less than 1%.

No other significant sources were detected in the imaging. The sensitivity limits of the imaging were determined by calculating the noise in concentric rings radiating out from the centroid position of the primary target, starting at one FWHM from the target with each ring stepped one FWHM from the previous ring. The 3σ limits of the J-band and K-band imaging were approximately 19 mag and 18 mag, respectively (see Figure 6). The J-band and K-band imaging limits are approximately 9 magnitudes fainter than the target which corresponds to approximately 10-11 magnitudes fainter than the target in the *Kepler* bandpass.

### *4.3. Speckle Imaging*

Speckle imaging of Kepler-22 was obtained on the night of 21 September 2010 UT using the two-color speckle camera at the WIYN 3.5-m telescope located on Kitt Peak. The speckle camera obtained 2000 30 msec EMCCD images simultaneously in two filters: R (692/40 nm) and I (880/50 nm). These data were reduced and processed to produce a final reconstructed speckle image for each filter. Figure 7 shows the reconstructed R and I band images. North is up and East is to the left in the image and the "cross" pattern seen in the image is an artifact of the reconstruction process. Seeing during the measurements was 0.86 arcsec. The details of the two-color EMCCD speckle camera and analysis procedure are presented in Horch et al. (2009), and Howell et al. (2011b).

For the speckle data, we determine if a companion star exists within the approximately 2.5 × 2.5 arcsec box centered on the target and robustly estimate the background limit in each summed, reconstructed speckle image. The two-color system allows us to have confidence in single fringe detection (finding and modeling identical fringes in both filters) if they exist and rule out companions between 0.05 arcsec and 1.5 arcsec from Kepler-22. We find no companion star within the speckle image separation detection limits to a magnitude limit of 4.24 mag in R (and 3.6  in mag in I), fainter than Kepler-22.

### *4.4. Search for a double-lined binary and nearby background stars*

Another type of search was conducted by taking spectra of the target star. If a background eclipsing binary is the source of a blend, it must be within 5.5 magnitudes of the target or it would be too faint to produce the observed transit amplitudes. The Keck-HIRES spectra should detect the lines of the offending background star in the spectrum of the target. Keck-HIRES spectra of Kepler-22 were used to search for lines of a background star. No such lines were observed.

In addition, a Keck spectrum was taken with the goal of detecting, or placing limits on, the contributions from any additional neighboring star located within 0.5 arcsec. The light from a closeby star, whether background or gravitationally bound to Kepler-22, would fall in the slit of the Keck spectrometer causing its spectrum to contaminate that of the main star, Kepler-22.

To detect any closeby star, we computed the cross-correlation function (CCF) for a large wavelength region of the Keck-HIRES echelle spectrum, from 360 - 620 nm. This region has few telluric lines, leaving the cross correlation dominated by stellar lines for FGKM stars. For the template we used a spectrum of Ganymede as a solar proxy. The cross-correlation function is very smooth, at the 1% level, from such a large wavelength range, and it nicely captures the shape of the thousands of absorption lines within that wavelength region.

We then cross-correlated a library spectrum (from a vast collection of 2000 Keck spectra of FGKM stars) of similar type, in this case HD 90156 (G5 V, $T_{\mathrm{eff}}$ = 5520 K). The goal is to compute the overall shape of the cross-correlation function from such stars, enabling a comparison of the CCF from Kepler-22 to the CCF obtained with this comparison star HD 90156. Indeed, the two cross-correlation functions came out with very similar shapes within 2% of the peak of the CCF. To detect the presence of any companion in Kepler-22, we took the difference between the CCF(Kepler-22) and CCF(HD 90156) to look for differences. Again, the differences were only typically 1-2% of the peak of the CCF, i.e. the two CCFs have nearly the same shape, as expected if there is no nearby star.

We then injected a second spectrum of a G star into the spectrum of the Kepler-22. This "fake" secondary spectrum can be adjusted to have any relative radial velocity and any relative intensity. We then compare the difference in CCFs, CCF(program star) - CCF(comparison), computed with and without the "fake" secondary spectrum. The fake secondary spectrum causes a "bump" in the CCF that departs from the CCF of the comparison star. Thus the difference between the two CCFs increases with increasing intensity of the fake secondary star.

We increased the relative intensity of the fake secondary star until the CCF of the fake binary system departs significantly from the CCF of the library comparison star. When the difference in CCF caused by the secondary star becomes larger than the systematic noise of the CCF, the secondary star is detectable.

Figure 8 is a representative plot of CCF(Kepler-22) - CCF(HD 90156) both as observed (dots) and also with a fake companion injected (solid line) in the spectrum of Kepler-22. For the case shown, the companion star has an intensity 0.025 of the primary star (i.e. 4 mag fainter), and it is separated by 30 km $s^{-1}$ (typically of either a background star or bound companion at 1 AU). This fake companion stands out against the systematic noise of the CCF and thus would be detectable. This case represents one example of a secondary spectrum that would have been detectable. Any companion 4 mag fainter (in the optical) and separated by at least 30 km $s^{-1}$ would have been detected at the $2\sigma$ level. At a separation of 10 km $s^{-1}$, the companion star would have to be brighter than delta-mag < 3 to remain detectable. No nearby stars were detected at these levels.

*4.5. Precise Doppler Measurements of Kepler-22*

We obtained sixteen high resolution spectra of Kepler-22 between 17 August 2010 and 25 August 2011 using the HIRES spectrometer on the Keck I 10-m telescope (Vogt et al. 1994) and four others; one at NOT (FIES), one at Fred Lawrence Whipple Observatory, and two at McDonald Observatory.

High-precision Doppler measurements are used to constrain the mass of Kepler-22b as discussed in Sections 5 and 6. (See Figure 9.) The uncertainty of the radial velocity measurements is based on the weighted uncertainty in the mean of the 700 spectral segments (0.2 nm long) that each contribute a separate Doppler measurement. The measured uncertainty is 1.4 m $s^{-1}$ per RV measurement. In addition, a proper assessment of precision should include an RV "jitter" of 3 m $s^{-1}$ for such stars, due to surface motions (Isaacson and Fischer 2010). The combination of the internal uncertainty and the jitter, added in quadrature, yields a final precision of ~4 m $s^{-1}$ for each measurement. (See Table 4.) Although the individual RV measurements have uncertainties of ~ 4 m $s^{-1}$, the MCMC analysis (Section 5) yields a posterior upper limit for the 1σ precision of the RV variation, constrained by the known period and ephemeris of the planet, of 4.9 + 6.7/ -7.4 m $s^{-1}$. The absence of a Doppler signal for Kepler-22b is used to compute an upper limit to the mass of this candidate under the planet interpretation.

*4.6. Warm Spitzer Observations*

Warm *Spitzer* observations in the near-infrared can also prove useful toward validating Kepler candidates, as shown for Kepler-10c (Fressin et al. 2011) and Kepler-19b (Ballard et al. 2011). The achromaticity of the transit depth, as compared between the optical *Kepler* photometry and near-infrared *Spitzer* photometry, provides an alternate means to confirm or reject the planetary nature of the candidate, since an eclipsing binary will present varying transit depths at different wavelengths unless the constituent stars have nearly identical colors (Torres et al. 2004, Tingley 2004).

We gathered observations using the Infrared Array Camera (IRAC) (Fazio et al. 2004) on Warm *Spitzer* at 4.5 μm the 1 October 2011 UT transit of Kepler-22b. These observations comprise part of a GO program (ID 80117, PI: D. Charbonneau) totaling 600 hours. The observations span 17 hours, centered on the 7.4-hour-long transit. We gathered the observations using the full-array mode of IRAC, with an integration time of 6 s/image. We employed the techniques described in Agol et al. (2010) for the treatment of the images before photometry. We first converted the Basic Calibrated Data products from the Spitzer IRAC pipeline (which applies corrections for dark current, flat field variations, and detector non-linearity) from mega-Janskys per steradian to data number per second. We identified cosmic rays by performing a pixel-by-pixel median filter, using a window of 10 frames. We replace pixels that are $> 4\sigma$ outliers within this window with the running median value. We also corrected for a striping artifact in some of the Warm Spitzer images, which occurred consistently in the same set of columns, by taking the median of the pixel values in the affected columns (using only rows without an overlying star) and normalizing this value to the median value of neighboring columns. Additionally, we remove the first hour of observations, while the star wandered to the position on the pixel where it spends the remaining hours of the observations.

We estimate the position of the star on the array using a flux-weighted sum of the position within a circular aperture of 3 pixels. We then performed aperture photometry on the images, using both estimates for the position and variable aperture sizes between 2.1 and 4.0 pixels, in increments of 0.1 pixels up to 2.7 pixels, and then at 3.0 and 4.0 pixels. We decided to use the position estimates using a flux-weighted sum at an aperture of 2.7 pixels, which minimized the out-of-transit RMS.

We remove the effect of the IRAC intrapixel sensitivity variations, or the "pixel-phase" effect (see eg. Charbonneau et al. 2005; Knutson et al. 2008) by assuming a polynomial functional form for the intrapixel sensitivity (which depends upon the $x$ and $y$ position of the star on the array). We denote the transit light curve $f$ (which depends upon time), and we hold all light curve parameters constant except for the transit depth. We use the light curve software of Mandel & Agol (2002) to generate the transit models. The model for the measured brightness $f'(x, y)$ is given by:

$$f' = f\left(t, \frac{R_p}{R_*}\right)\left(b_1 + b_2(x-\bar{x}) + b_3(x-\bar{x})^2 + b_4(y-\bar{y}) + b_5(y-\bar{y})^2\right) \qquad \text{Eq. 1}$$

where we include all of the observations (both in- and out-of-transit) to fit the polynomial coefficients and the transit depth simultaneously.

We fit for the polynomial coefficients $b_1$ through $b_5$ using a Levenberg-Marquardt $\chi^2$ minimization. The *Spitzer* light curve contains significant correlated noise even after the best intrapixel sensitivity model is removed. We incorporate the effect of remaining correlated noise with a residual permutation analysis of the errors as described by Winn et al. (2008), wherein we find the best-fit model *f'* to the light curve as given by Equation 1, subtract this model from the light curve, shift the residuals by one step in time, add the same model back to the residuals, and refit the depth and pixel sensitivity coefficients. We wrap residuals from the end of the light curve to the beginning, and in this way we cycle through every residual permutation of the data. We determine the best value from the median of this distribution, and estimate the error from the closest 68% of values to the median. We gathered 8.4 hours of observations outside transit, which is sufficient to sample the systematics on the same timescale as the 7.4 hour transit. Using the residual permutation method on the light curve treated with a polynomial, we find a best-fit transit $R_p/R_\star$ to be 0.0184 ± 0.0050, a 3.7σ detection.

We note that we also treated the light curve with the weighted sensitivity function used in Ballard et al. (2010), which proved in that work to produce a time series with lower RMS residuals. For this procedure, we do not assume any a priori functional form for the intrapixel sensitivity; rather, we perform a weighted sum over neighboring points for each flux measurement, and use this sum to correct each flux measurement individually. However, during these observations we observed an added component of pointing drift in the X direction, comparable to the drift in the Y direction of 0.1 pixels. This drift resulted in few out-of-transit observations that overlap on the pixel with in-transit observations. We found no improvement using the weighted sensitivity method (which depends strongly upon the existence of out-of-transit observations to model the portion of the pixel at which the transit occurs), as compared to the polynomial method—while the best-fit transit depth was similar, the error bars were 20% larger using the former method.

This value of the transit depth measured with Warm *Spitzer* of 340 ± 200-160 ppm is in agreement with the depth in the *Kepler* bandpass of 492 ± 10 ppm (corrected for limb-darkening), which favors the planetary interpretation of the light curve. In Figure 10, we show the binned *Spitzer* light curve (by a factor of 300), with the best-fit transit model derived from the *Spitzer* observations and the best-fit *Kepler* transit model (corrected for limb-darkening) overplotted. We comment further on the types of blends we rule out by BLENDER in Section 6.

## 5. DETERMINATION OF STELLAR PARAMETERS

### *5.1. Reconnaissance Spectroscopy*

Spectroscopic observations and analysis to determine the stellar characteristics of Kepler-22 were conducted independently at several observatories. After preliminary vetting and recognizing the importance of this candidate, further analysis was conducted. LTE spectroscopic analysis using the spectral synthesis package SME (Valenti & Piskunov 1996; Valenti & Fischer 2005) was applied to a high resolution template spectrum from Keck-HIRES to derive an effective temperature, $T_{\rm eff}$ = 5518 ± 44 K, surface gravity, log $g$ = 4.44 ± 0.06 (cgs), metallicity, [Fe/H] = −0.29 ± 0.06, $v \sin i$ = 0.6 ±1.0 km $s^{-1}$, and the associated error distribution for each of them. To refine the true parameters of the star, we used these observations of the effective temperature to constrain the fundamental stellar parameters derived via asteroseismic analysis.

As an independent check of the values of the SME parameters, we also derived values by matching the spectrum to synthetic spectra (Torres et al. 2002, Buchhave et al. 2010), and in addition we employ a new fitting scheme that allows us to extract precise stellar parameters from the spectra. We report the mean of the spectroscopic classification of one HIRES spectrum, one spectrum from the fiber-fed Échelle Spectrograph (FIES) on the 2.5 m Nordic Optical Telescope (NOT) on La Palma, Spain, one spectrum from the fiber-fed Tillinghast Reflector Échelle Spectrograph (TRES) on the 1.5 m Tillinghast Reflector at the Fred Lawrence Whipple Observatory on Mt. Hopkins, Arizona and two spectra from the Tull Coudé Spectrograph on the 2.7 m the Harlan J. Smith Telescope at the McDonald Observatory Texas acquired between August 2009 and July 2011. The analysis yields $T_{\rm eff}$ = 5642 ± 50 K, log $g$ = 4.49 ± 0.10, [m/H] = -0.27 ± 0.08, $v$ sin $i$ = 2.08 ± 0.50 km $s^{-1}$, which agrees with the results from the SME analysis of the HIRES spectrum within the uncertainties.

### *5.2. Ca II H& K activity*

Using the high resolution spectra acquired with HIRES at Keck Observatory, we have monitored the chromospheric emission via the Ca II H&K lines. These lines are used to monitor stellar activity, magnetic variability and rotation rates for main sequence stars (Baliunas 1995, Noyes 1984). Using stars observed at both Mt. Wilson and Keck observatories, Isaacson & Fischer(2010) calibrated the Ca II H&K flux measurements from Keck to the Mt. Wilson activity scale. The ratio of the flux in the cores of the Ca II H&K lines relative to the continuum flux yields a Mt Wilson S-value equal to 0.149+-0.004. The S-value is parameterized as log $R'_{HK}$, the fraction of flux in the cores of the H&K lines compared to the total bolometric emission. Using log $R'_{HK}$ allows for comparison of stellar activity for different stellar types regardless of continuum flux near the Ca II H&K lines. The measured log $R'_{HK}$ of -5.087 ±0.05 indicate that the star is inactive, which is consistent with the slow rotation rate found spectroscopically. These results imply that Kepler-22 is an old star.

*5.3. Asteroseismic Observations*

The data series for Kepler-22b contains 19 months of data taken at a cadence of 1 minute during *Kepler*-observing quarters Q2 (month 3) through Q8 (month 3; 24 August 2010 to 22 September 2010). At Kp = 11.664, the star is relatively dim, which makes detection of signatures of solar-like (p-mode) oscillations a challenging task because oscillation amplitudes are expected to be below the solar level.

The power spectrum of the lightcurve does not show a clear excess of power. However, based on asteroseismic analysis of the data using the pipeline developed at the Kepler Asteroseismic Science Operations Center (as described in detail by Christensen-Dalsgaard et al. (2008, 2010); Huber et al. (2009); Gilliland et al. (2011)) a p-mode signal can be detected and extracted. See Figure 11. We used a matched filter approach to search for and determine a value for the average large frequency separation of the oscillations spectrum, as well as a frequency for the maximum p-mode power. The p-mode signal is located near 3.15 mHz and has a peak amplitude of approximately 3.4 ppm for radial modes. Several search algorithms were used to estimate the large separation, details of which may be found in Hekker et al. (2010) and Verner & Roxburgh (2011). The large frequency separation for p-modes, which is the prime average asteroseismic parameter (see e.g. Gilliland et al., 2011 for details), was determined to be 137.5 ± 1.4 microHz, which is only 1.9 ± 1.0 per cent larger than the solar large frequency separation. This value for the large separation indicates that the mean stellar density for Kepler-22 is 3-4 % larger than the solar mean density.

Using a series of stellar models that fit the large frequency separation, combinied with observed properties for metallicity and effective temperature, the stellar radius, mass, chemical composition and the effective temperature are inferred (Stello et al., 2009, Basu et al., 2010, Metcalfe et al., 2009, Christensen-Dalsgaard et al., 2010, Quirion, et al. 2010, Gai et al. 2011). The uncertainties on the estimated stellar properties provided by these fits are based on error propagation through the series of stellar models. The accepted values for the stellar surface gravity, density, mass, radius, and luminosity were derived from this analysis and are listed in Table 1.

As part of the analysis we also searched for individual p-mode frequencies which fit the detected excess power and frequency pattern. The aim of this search was to constrain the small frequency separation which would in principle allow us to estimate the age of the star and the core Helium content. Although we find frequencies that fit the expected p-mode structure, we consider the detection to be too weak to perform a detailed modeling of individual frequencies. The risk of performing a detailed frequency modeling on a weak signal is that this could provide misleading conclusions on the system age. Therefore no age is shown in Table 1.

## 6. ESTIMATION OF THE PROBABILITY OF A FALSE-POSITIVE EVENT

In this section we examine the possibility that the transit signals seen in the *Kepler* photometry of Kepler-22b are the result of contamination of the light of the target by an eclipsing binary (EB) along the same line of sight ("blend"). The eclipsing binary may be either in the background or foreground, or at the same distance as the target in a physically associated configuration (hierarchical triple). The object eclipsing the intruding star can be either another star or a planet.

We explore the wide variety of possible false positive scenarios using the BLENDER technique (Torres et al. 2004, Torres et al.2011; Fressin et al. 2011), a procedure that allows the validation of Kepler-22b independently of the detection of the reflex motion of the star (radial velocities). BLENDER generates synthetic light curves for a large number of blend configurations and compares them with the Kepler photometry in a chi-square sense. The parameters considered for these blends include the masses of the two eclipsing objects (or the size of the eclipser, if a planet), the relative distance between the binary and the target, the impact parameter of the transiting object, and the eccentricity and orientation of the orbit of the eclipsing binary, which can affect the duration of the events. These parameters are varied in a grid pattern over broad ranges. Scenarios that give fits significantly worse than a planet model fit (at the 3-$\sigma$ level) are considered to be rejected. While this reduces the space of parameters for viable blends considerably, it does not eliminate all possible blends. Constraints from follow-up observations described previously (such as high-resolution imaging and spectroscopy) as well as multi-band photometry available for the target are then used to rule out additional parts of parameter space. The frequency of false positive scenarios that remain after these efforts is assessed statistically, in the manner we describe below. Adopting a Bayesian approach, this blend likelihood is then compared with an a priori estimate of the likelihood of a true planet (odds ratio). We consider the candidate to be statistically "validated" if the likelihood of a planet is much greater (several orders of magnitude) than that of a blend. Examples of other *Kepler* candidates validated in this way include Kepler-9d (Torres et al. 2011), Kepler-10c (Fressin et al.

2011), Kepler-11g (Lissauer et al. 2011a), Kepler-18b (Cochran et al. 2011), and Kepler-19b (Ballard et al. 2011).

Illustration of the constraints on false positives provided by BLENDER for Kepler-22b are shown in Figure 12 – 14: first for blends involving a background eclipsing binary composed of two stars, then for background or foreground stars transited by a larger planet and finally for cases of hierarchical triple systems with a secondary transited by a planet. Following the BLENDER nomenclature we refer to the target star as the "primary", and to the components of the eclipsing binary as the "secondary" and "tertiary". The space of parameters in Figure 12 is projected along two of the dimensions, corresponding to the mass of the secondary and to the relative distance between the primary and the binary (cast for convenience here in terms of the difference in distance modulus in magnitudes). The colored regions represent contours of equal goodness of fit compared to a transiting planet model, with the 3-sigma contour indicated in white. Blends inside this contour give acceptable fits to the *Kepler* photometry, and are considered viable. They all involve eclipsing binaries that are up to ~ 5.5 magnitudes fainter than the target (dashed green line in the figure). Other constraints can potentially rule out additional blends. For example, blends in the blue-hatched areas have overall colors for the combined light that are either too red (left) or too blue (right) compared to the measured color of the target ($r-K_s = 1.475 \pm 0.022$, taken from the KIC; Brown et al. 2011), at the $3\sigma$ level. For this particular kind of blend these constraints are not helpful however, as those scenarios are already ruled out by BLENDER. False positives that are in the green-hatched area correspond to secondary components that are less than one magnitude fainter than the target, and which we consider to be also ruled out because such stars would usually have been detected in our spectroscopic observations, as a second set of lines. Once again this constraint is redundant with the BLENDER results. The one-mag limit is very conservative, as stars down to 2 or 3 magnitudes fainter than the target would also most likely have been seen in our high-resolution, high signal-to-noise ratio Keck spectra.

A similar diagram for blends involving background or foreground stars orbited by a transiting planet is presented in Figure 13. In this case both the color index constraint and the brightness constraint significantly reduce the space of parameters in which blends can reside, which is indicated by the thick white contour ("Allowed Region"). Within this area only tertiaries that are between 0.32 $R_{Jup}$ and 2.0 $R_{Jup}$ in size are able to produce signals that are consistent with the observations. These false positives are all in the background, and can be up to 5 magnitudes fainter than the target in the *Kepler* bandpass, as indicated by the dashed green line.

BLENDER easily rules out all hierarchical triple configurations with stellar tertiaries, as these invariably lead to the wrong shape for a transit. However, planetary tertiaries of the right size can still mimic the light curve well. The landscape for this type of blend is seen in Figure 14. For Kepler-22b the combination of the color and brightness constraints allows us to reject all hierarchical scenarios.

### *6.1 Validation of Kepler-22b*

The tight restrictions on blends that are able to match the detailed shape of the transit allow us to estimate the expected frequency of these scenarios. We follow a procedure analogous to that described by Fressin et al. (2011). For blends with stellar tertiaries, this frequency will depend on the density of background stars near the target, the area around the target within which such stars would go undetected, and the rate of occurrence of eclipsing binaries. We perform these calculations in half-magnitude bins, with the following inputs: a) the Galactic structure models by Robin et al. (2003) to estimate the number of stars per square degree, subject to the mass limits allowed by BLENDER; b) results from our adaptive optics observations to estimate the maximum angular separation ($\rho_{max}$) at which companions would be missed, as a function of magnitude difference relative to the target ($K_p$ = 11.664); and c) the overall frequency of eclipsing binaries capable of mimicking the transits (0.78%; Fressin et al. 2011). Table 5 presents the results. Columns 1 and 2 give the magnitude range for background stars and the magnitude difference compared to the target; columns 3 and 4 list the mean star densities and $\rho_{max}$, and column 5 (the number of background stars we cannot detect) is the result of multiplying column 3 by the area implied by $\rho_{max}$. Finally, the product of column 5 and the eclipsing binary frequency of 0.78% leads to the blend frequencies in column 6. The sum of these frequencies is given at the bottom, under "Totals".

Similar calculations are performed for scenarios in which the tertiaries are planets instead of stars, and the results are presented in columns 7-10 of the table. The planet frequencies adopted for this calculation have been taken from the census of candidates detected by *Kepler*, described below. Adding up the contributions from the two types of blends ($0.977 \times 10^{-6}$ for stellar tertiaries, and $0.184 \times 10^{-6}$ for planetary tertiaries), we obtain a total blend frequency of BF = $1.2 \times 10^{-6}$.

We next require an estimate of the expected frequency ($P_f$) of true transiting planets similar to Kepler-22b ("planet prior[4]"), to assess whether the likelihood of a blend is sufficiently smaller than that of a planet in order to validate Kepler-22b. A reasonable estimate for PF may be obtained by examining the list of candidates from the *Kepler Mission* itself, which currently contains over 1235 candidates (Borucki et al. 2011) detected using observations gathered from Q1 to Q6. These candidates have been subjected to various levels of vetting, including at least the following: checking for the presence of secondary eclipses that might betray a blend, making sure that the odd- and even-numbered events are of the same depth, looking for consistency in the transit depth from quarter to quarter, verifying that the shape in each quarter is transit-like, and examining the flux centroids to rule out displacements that might be due to a blended star in the photometric aperture. While these candidates have not yet been confirmed because follow-up observations are still in progress, the false positive rate is expected to be relatively small (typically less than 10%; see Morton & Johnson 2011), so for our purposes we have assumed that all of them represent true planets. In this sample there are 437 cases that have planetary radii within 3σ of the measured value for Kepler-22b ($R_p = 2.35 \pm 0.12\ R_\oplus$), where we have used this 3σ limit for consistency with a similar criterion adopted above in BLENDER. (No constraint is placed on the orbital period in either case.) Considering the total number of 190,186 *Kepler* targets examined between Q1 and Q6, we obtain a planet frequency ($P_f$) of 437/190,186 = 2.3 x $10^{-3}$, which is significantly larger than the blend frequency ($B_f$), by a factor of about 1900.

It may be argued, however, that the above calculation of both $B_f$ and $P_f$ should be restricted to planets of similar orbital period as Kepler-22b, which is fairly long (290 days), as the rate of occurrence may be different for short-period and long-period planets, and this could alter the odds ratio. If, for example, we limit the periods to be within a factor of two of 290 days, we find that the blend and planet likelihoods are both significantly smaller, and that they indeed change by different amounts. $B_f$ is reduced to 2.9 x $10^{-8}$ (stellar tertiaries) + 1.6 x $10^{-8}$ (planetary tertiaries) = 4.5 x $10^{-8}$, and $P_f$ is reduced to = 5/190,186 = 2.6 x $10^{-5}$. (The numerator does not include the candidate itself.) The odds ratio then becomes $P_f/B_f$ = 578, which is still large enough that it allows us to validate Kepler-22b with a very high degree of confidence.

[4] "planet prior" is the *a priori* probability that detected event is caused by a transiting planet.

Furthermore, we consider the above odds ratio of ~600 to be a conservative estimate in the sense that it does not include corrections for the fact that shallow transits such as those of Kepler-22b can only be detected in a fraction of the 190,186 *Kepler* targets. Incompleteness may affect the blend frequencies as well, but will do so to a much smaller degree because the planets involved in blends are larger (0.32 to 2.0 $R_{Jup}$) and have deeper transits that are easier to detect. Thus, we consider the planet prior adopted above to be a conservative estimate, which strengthens our conclusion on the true planetary nature of Kepler-22b.

## 7. MODEL ANALYSIS TO DETERMINE PLANET CHARACTERISTICS

Based on the analysis of stellar spectra observations and the asteroseismic analysis described in the previous sections, the planetary radius is determined with a precision of just over 5%. However, estimates of the planet mass are driven by the precision of the radial velocities and their distribution in phase. A Markov Chain Monte Carlo (MCMC) model analysis was used to derive estimates for the mass and other planetary parameters.

We adopted simple aperture photometry from the Kepler pipeline for our transit analysis. This data include pixel corrections such as smear and background levels. At this level of correction, the photometry still includes differential instrumental effects and astrophysical variability such as spot modulation. The aperture photometry was detrended using a second-order polynomial that is fit to continuous segments of Kepler data. A segment is defined as a set of observations that are uninterrupted for less than 5 cadences (~2.5 hours). Observations that occurred during transit where masked off when calculating the best fit. The best fit polynomial was removed from each segment and the entire lightcurve was normalized by the median.

The photometric and radial velocity measurements were fit to a model to measure the physical and orbital properties of the star and companion. The model fits for the means stellar density ($\rho$*), period (P), epoch (T0), impact parameter (b), the scaled planetary radius ($R_p/R*$), eccentricity and argument of pericenter ($e \cos \omega$, $e \sin \omega$), radial velocity amplitude (K), and the radial velocity zero point (gamma). Due to the long period, there was no need to account for the effects of reflected and emitted light from the planet, ellipsoidal variations due to tidal distortions of the host star, and Doppler boosting due to motion of the star around the center of mass. The transit shape is described by the analytic formulae of Mandel & Agol (2002) and the planet orbit is assumed to Keplerian. We use the fourth-order non-linear parameterization of limb-darkening with coefficients (c1 = 0.4599, c2 = 0.1219, c3 = 0.4468, c4 = -0.2800) as calculated by Claret and Bloem (2011) for the *Kepler* bandpass. We first computed a best fit model by minimizing chi-squared using a Levenberg-Marquardt algorithm. The asteroseismic constraint on $\rho$* was used as a prior.

The best-fit model was then used as a seed for a hybrid MCMC routine to determine posterior distributions of our model parameters. The model is considered a hybrid, as we randomly use a Gibbs-sampler and a buffer of previous chain parameters to produce proposals to jump to a new location in the parameter space. The addition of the buffer allows for a calculation of vectorized jumps that allow for efficient sampling of highly correlated parameter space.

The posterior distributions of the stellar mass and radius as determined by asteroseismology are convolved with the posterior distributions from our model parameters to compute the planetary mass, radius, orbital inclination and semi-major axis. Model results for the best fit to the transit pattern give a relative transit depth of 491.9 +9.1/-10.9 ppm dimming lasting 7.415 +0.067/-0.078 hours with a transit ephemeris of T[BJD] = 245966.6983 ±0.0023 and period of 289.8623 +0.0016/-0.0020 days.

The median of the distribution for each model parameter and the corresponding ±68.3% credible intervals (akin to 1σ confidence interval) centered on the median are tabulated in Table 1. As our models allow for fully eccentric models, the true upper limits on the mass of the planet can be estimated. For 1, 2, and 3 σ, the upper limits are 36, 82 and 124 $M_\oplus$. For a circular orbit, the upper limits on the mass are 27, 50 and 71 $M_\oplus$.

## 8. SEARCH FOR TRANSIT TIMING VARIATIONS

Kepler-22b has high SNR transits and a long orbital period; transit timings of such planets are quite sensitive to perturbations by other planets in the system (Holman et al 2005,

Agol et al 2005). The individual transit times are listed in Table 6. The three values had residuals from a constant-period model of 0.7 ± 3.1 min, -1.1 ± 2.7 min, and 0.6 ± 2.8 min, respectively. The individual transit times are quite consistent with a constant period, but to quantify the possible variation, we resampled mid-transit times by adding Gaussian perturbations of σ equal to the quoted error bar, and fit out a linear ephemeris, in $10^4$ trials. The resulting distribution of the RMS value of the transit timing variation had a mode near 0 minutes and a 2σ upper limit of 3.5 minutes, or $8\times10^{-6}$ of an orbital period.

We wish to compare this upper limit to the timing variations expected from the class of super-Earth and Neptune planets (SEN, defined as $M_p \sin i < 30\ M_\oplus$. The Doppler survey HARPS (Mayor et al. 2011, Table 3) has reported 63 such planets, among which all but 8 are in multiple-planet systems. We performed numerical integrations of the 33 systems containing the remaining 55 SEN planets, and extracted the timing signals for 102 orbits via the method of Fabrycky et al. (2010). The orbital planes were assumed to be coplanar and edge-on; planetary masses were chosen as their measured minimum masses; orbital phase (mean anomaly λ and direction of periastron (ω) were drawn uniformly from 0 to 2 π. From these integrations 100 sets of three adjacent transits were used to compute the distribution of the RMS value of the transit timing variation for each planet. These distributions for all 55 planets were combined to form the histogram shown in Figure 15. The peak value implies that a typical TTV value for SEN planets is $\sim10^{-5}$ planetary orbital periods: that corresponds to ~ 4 minutes for Kepler-22b.

Another relevant comparison of a habitable-zone planet is the timing variations the Earth experiences, as would be measured in transit from afar (Holman et al. 2005). We computed its timing signature over 1000 orbits, in the presence of the 7 other planets, and computed the RMS value of the transit timing variation for each set of three adjacent transits; that distribution is also plotted on Figure 15.

Our limit to the timing variations of Kepler-22b is close to or below the values for many known systems. We are thus sensitive to the transit-timing signature of low-mass planets in the habitable zone. This fact will become important over time, as we seek to confirm and measure the masses of such planets. The unique interpretation of such timing deviations, however, requires an extended mission. For instance, the mass measurements in Kepler-11 used between 9 and 41 transits for each planet for confirmation (Lissauer et al. 2011b). A mission duration of 8 years would enable similar studies on systems with planets in the habitable zone.

## 9. HABITABLE ZONE DISCUSSION

### *9.1. Composition of Kepler-22b*

Because only an upper limit to the mass of Kepler-22b is available (36 $M_\oplus$,1σ), any density less than 14.7 g/cc is consistent with the observations; i.e., the composition is unconstrained. Several planets with sizes similar or less than that of Kepler-22b have been discovered that have densities too low for a rocky composition (Lissauer et al. 2011). However, others, such as Kepler-18b have a size ($R_p$ = 1.98 $R_\oplus$) similar to Kepler 22b and a density (4.9 ± 2.4g/cc) sufficiently high to imply that such planets could have a solid or liquid surface. Further, model studies of planetary structure often consider rocky planets with masses of 100 $M_\oplus$ or more (Ida and Lin, 2004, Fortney et al., 2007). Because there is a possibility that Kepler-22b is a planet with a surface and an atmosphere, a surface temperature will be estimated.

*9.2. Estimated surface temperature for a rocky planet at the distance of Kepler-22b from its host star*

The habitable zone (HZ) is often defined to be that region around a star where a rocky planet could have a surface temperature between the freezing point and boiling point of water, or the region receiving the same insolation as the Earth from the Sun (Rampino and Caldeira 1994, Kasting 1993, Heath et al. 1999, Joshi 2003, Tarter et al, 2007).

The radiative equilibrium temperature for a planet is estimated from:

$$T_{\mathrm{eq}} = \left(\frac{(1-\alpha)(R_*)^2}{4\beta a^2}\right)^{1/4} T_{\mathrm{eff}} = 262\ \mathrm{K} \qquad \text{Eq.2}$$

where $T_{\mathrm{eff}}$ is the effective temperature of the star (5518 K), $R_*$ is the radius of the star, α is the planet Bond albedo, *a* is the planet semi-major axis, β represents the fraction of the surface of the planet that reradiates the absorbed flux (assumed to be 1.0 for a rapidly rotating body with a strongly advecting atmosphere), and $T_{\mathrm{eq}}$ is the radiative equilibrium temperature of the planet. The calculations assume a Bond albedo equal to that of the Earth (0.29). The uncertainty in the computed equilibrium temperature is approximately 22% because of uncertainties in the stellar size, mass, and temperature as well as the planetary albedo, but almost entirely due to the latter.

$T_{\mathrm{eq}}$ is the temperature at which the insolation balances the thermal radiation from the planet. It is equal to the surface temperature $T_s$ only for a planet lacking an atmosphere. A planet with an atmosphere will have a surface temperature above $T_{\mathrm{eq}}$ because of the warming caused by the atmosphere. For example, the greenhouse effect raises the Earth's surface temperature by 33

K and that of Venus by approximately 500 K. Further, the spectral characteristics of the stellar flux vary strongly with $T_{eff}$. This factor affects both the atmospheric composition and the chemistry of photosynthesis (Kasting et al. 1993, Heath et al 1999, Segura et al 2005). Using Equation 2, and assuming a planet with a surface and an atmosphere with thermal properties similar to that of the Earth (which is unlikely) and a Bond albedo of 0.29, the surface temperature of Kepler-22b would be approximately 295 K.

## 10. SUMMARY

Based on the three transits observed in the 22-month period between 12 May 2009 and 14 March 2011, a planet with an orbital period of 289.8623+0.0016/-0.0020 days and with a relative transit depth of 492 ±10 ppm and a transit duration of 7.415+0.067/-0.078 hours has been validated. High spatial-resolution images show no evidence for any companion star near enough to affect the light curve of this system. The precision of the radial velocity measurements is not sufficient to determine a mass, but provide a 1σ upper limit of 36 $M_{\oplus}$ derived from a combination of RV observations and modeling. After eliminating all scenarios that are not consistent with the data, the results indicate a planet with a semi-major axis of 0.840±0.012 AU and a radius of 2.38±0.13 $R_{\oplus}$. Based on the host star's temperature, size, and mass, the calculated $T_{eq}$ is 262 K, similar to 255 K for the Earth (Allen, 1999). In the event that this planet has a surface and an atmosphere that provides a modest amount of greenhouse heating, the surface temperature would be appropriate for liquid water to exist on the planet's surface. This places it in the habitable zone (HZ) of Kepler-22. Radial velocity surveys have found many giant planets in circumstellar habitable zones, and a few moderate-mass planets in or near the HZ have been discovered recently (Pepe et al. 2011). Kepler 22b is the smallest planet with known radius, in the HZ of any star other than the Sun.

## Acknowledgments

*Kepler* was competitively selected as the tenth Discovery mission. Funding for this mission is provided by NASA's Science Mission Directorate. Some of the data presented herein were obtained at the W. M. Keck Observatory, which is operated as a scientific partnership among the California Institute of Technology, the University of California, and the National Aeronautics and Space Administration. The Keck Observatory was made possible by the generous financial support of the W. M. Keck Foundation. Some of the observations were made with the Spitzer Space Telescope, which is operated by the Jet Propulsion Laboratory, California Institute of Technology under a contract with NASA. Support for this work was provided by NASA through an award issued by JPL/Caltech. Additional support for this work was also received from National Center for Atmospheric Research which is sponsored by the National Science Foundation. The authors would like to thank the many people, including William Rapin who gave so generously of their time to make this Mission a success. SH acknowledges financial support from the Netherlands Organisation for Scientific Research (NWO).

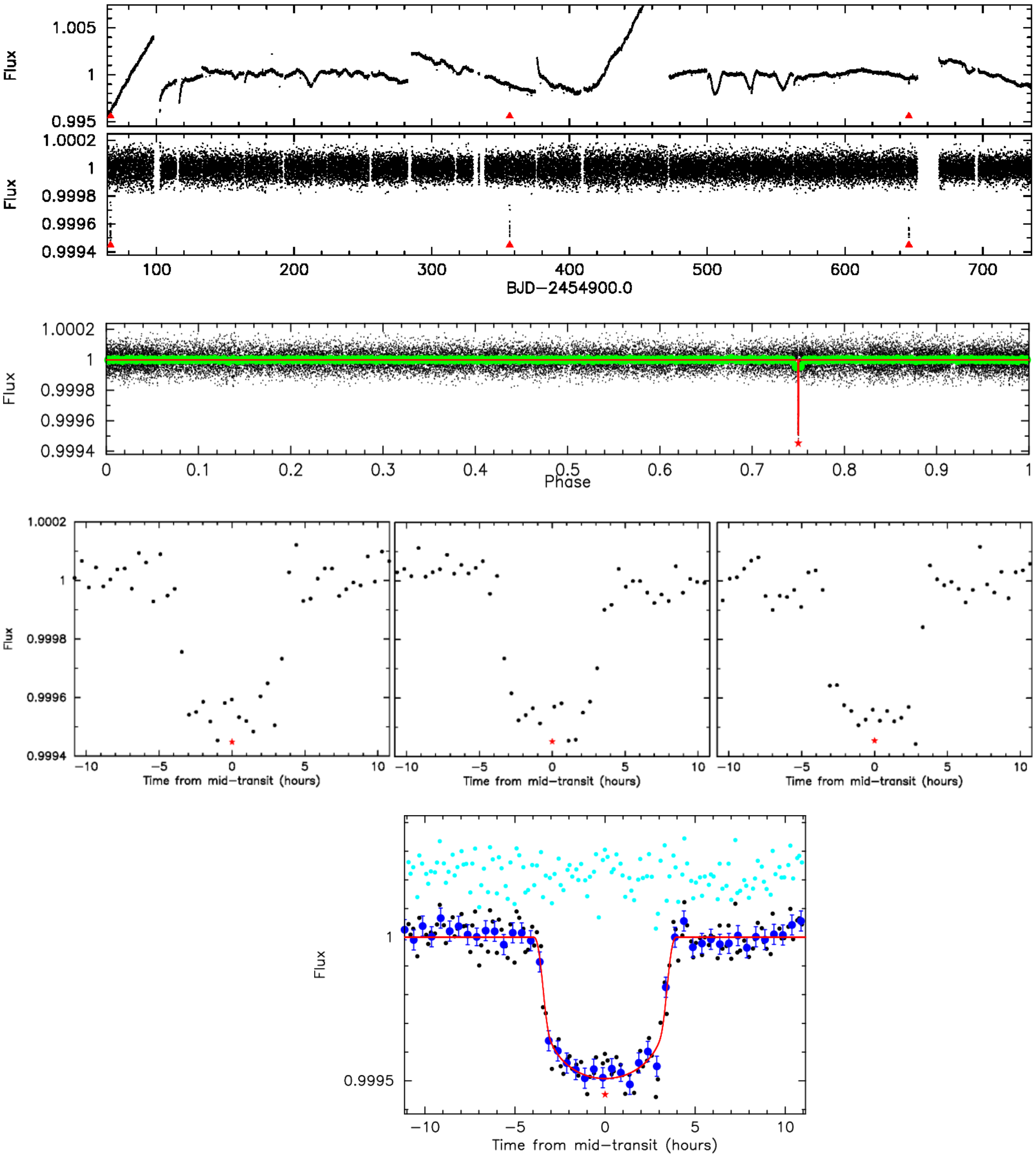

Figure 1. Light curves for Kepler-22b. Top panel: raw aperture flux time series. Second panel: flux time series after removal of a second-order polynomial for each segment and normalizing the data of each quarter by the median. No occultation is detected at any phase. Red triangles mark the location of each transit, Third panel: Phased light curve. Black dots show the long-cadence data, the green dots show the data averaged into 100 evenly spaced bins in phase. The red line shows the best bit model. Fourth panel: Individual transits show same depth and width consistent with a planetary transit. Fifth panel: folded light curve with model fit in red. Black dots represent individual observations. Dark blue points represent 30-minute binned data, and cyan points represent residuals after fitting. Red asterisk represents the mid-transit times based on the model fit with eccentricity value allowed to float.

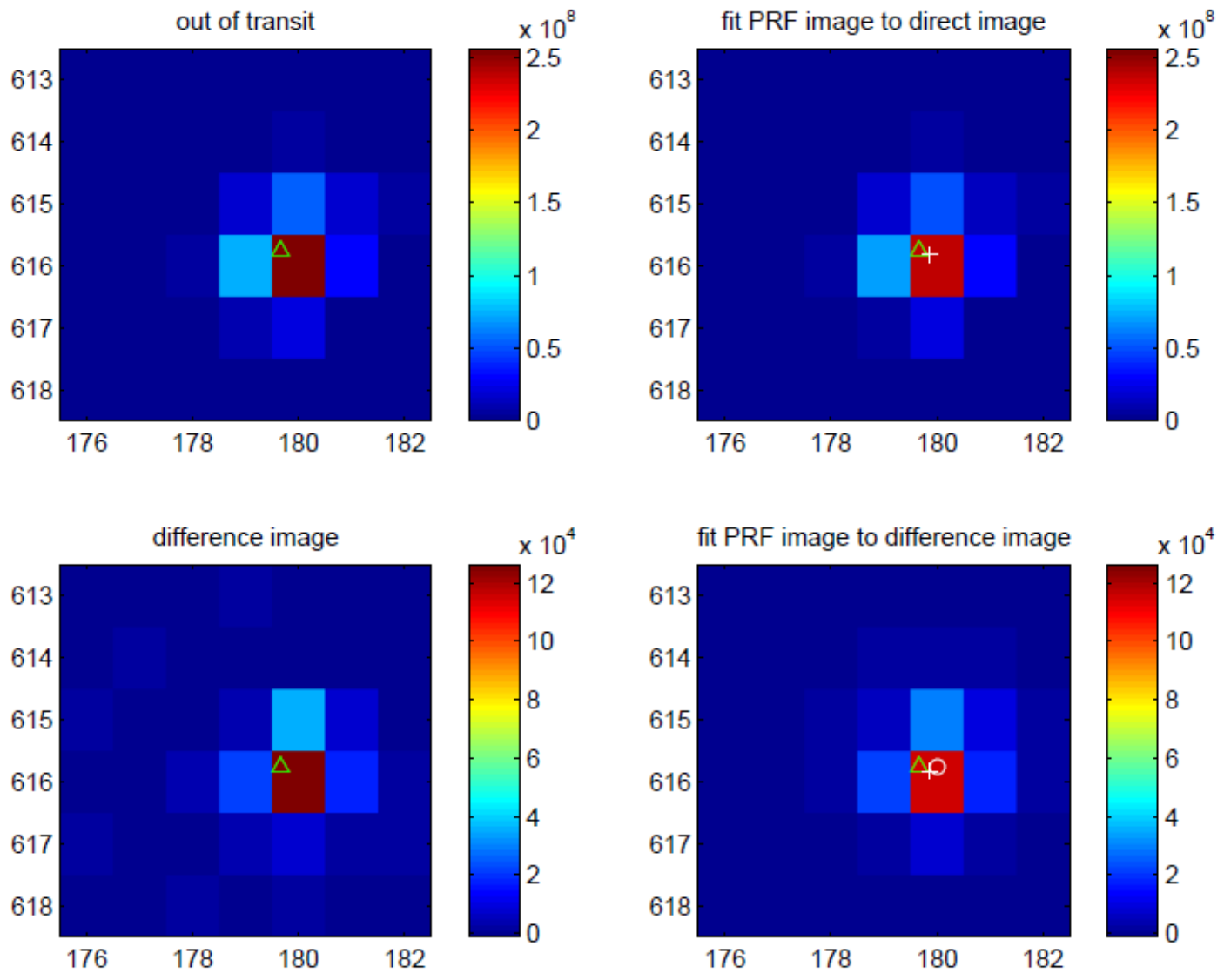

Figure 2. Kepler pixels from Q1, showing average out-of-transit (top left) and difference (bottom left) images. Pixels reconstructed using the PRF fit to the out of transit (top right) and difference image (bottom right) are shown for comparison, indicating that the fitted PRFs match the data. Symbols show the location of the fitted PRFs relative to the catalog position of the star. Green △: catalog position of the target. White +: PRF fit to the direct image. White ○: PRF fit to difference image.

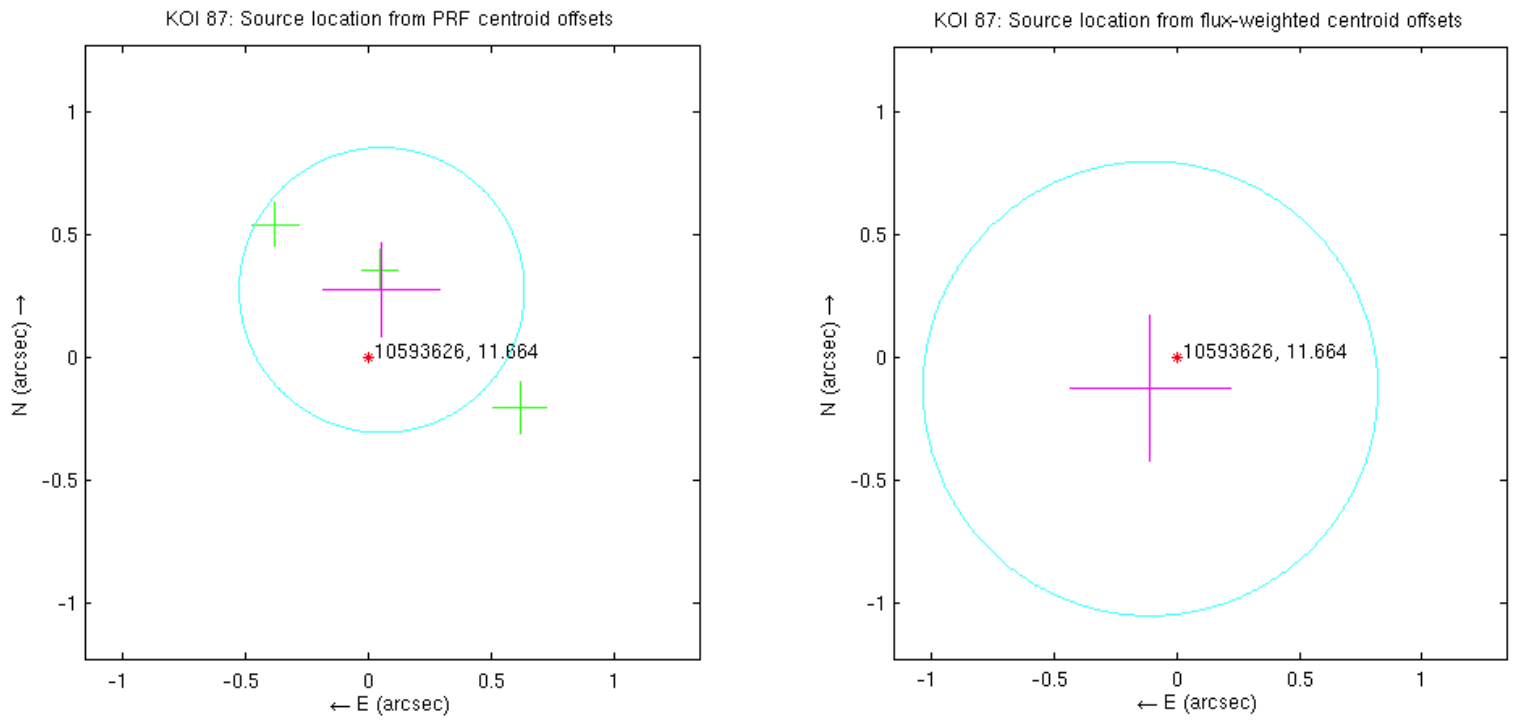

Figure 3. Multi-quarter centroid analysis offset results in coordinates centered on Kepler-22. Left: PRF-fit where the green crosses show the fit to the individual observed transits, and the magenta cross shows their average positions. The length of the crosses is the 1σ uncertainty in RA and Dec, and the cyan circle is the 3σ uncertainty around the average observed offset. The flux-weighted figure (right) shows only the multi-quarter result with the same symbols.

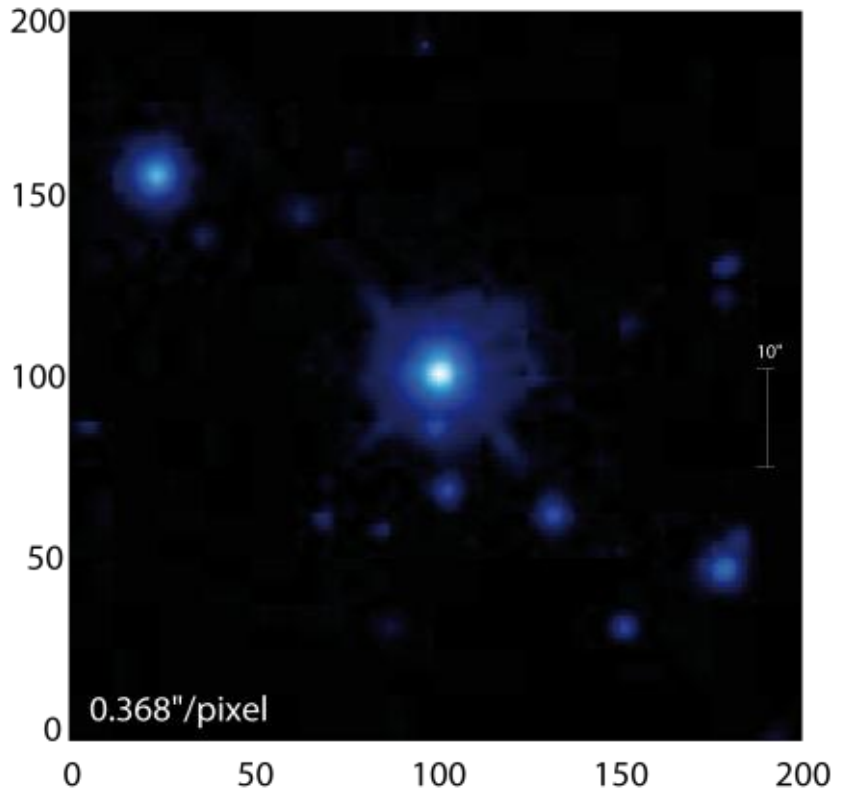

Figure 4. Image of the star field near Kepler-22. I-band image; N is up. E is left. Field of view is 1.2’x 1.2’. Kepler-22 is at the center of the image.

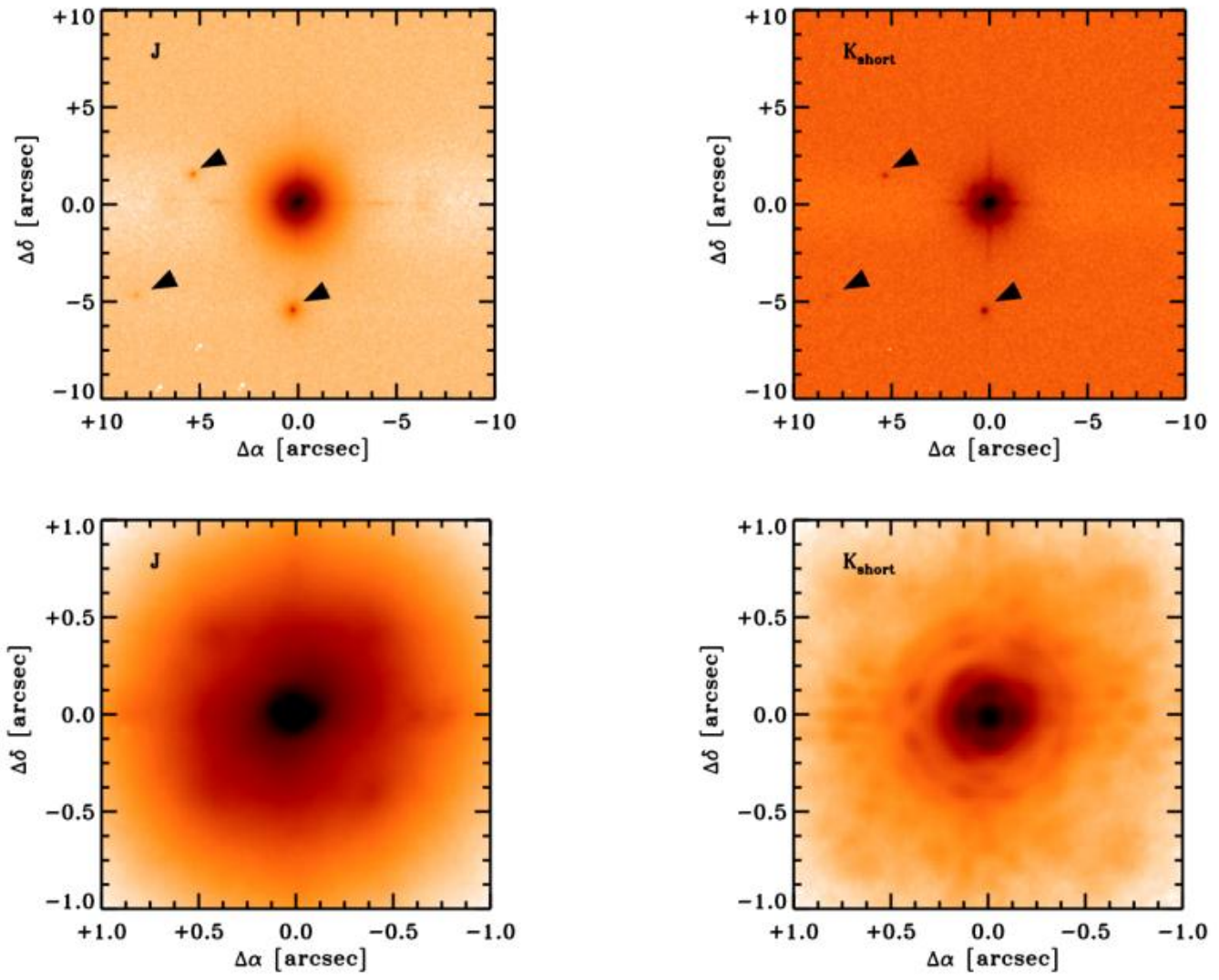

Figure 5: J and Ks Palomar adaptive optics images of Kepler-22b. The top row displays a 20″ × 20″ field of view (FOV) centered on the primary target. The bottom row displays the central 2″ ×2″ FOV. Vertical and horizontal arrows mark nearby stars to the East and South, respectively. North is up and East is to the left.

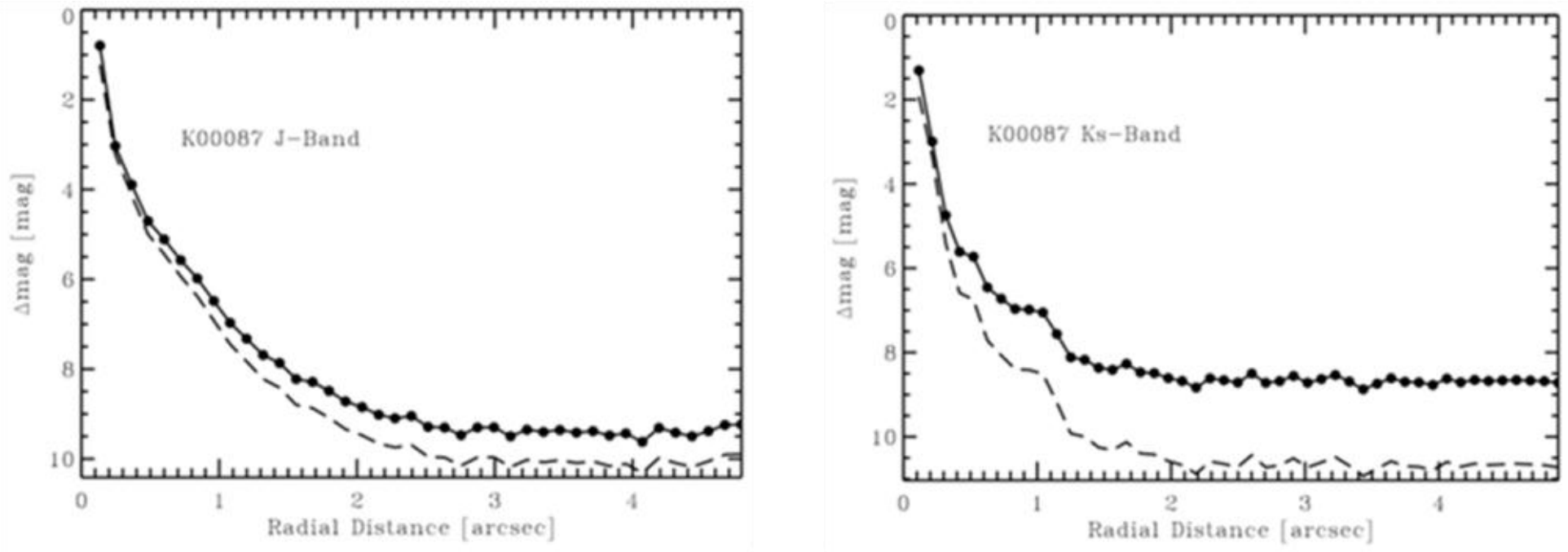

Figure 6**.** Left: The sensitivity limits of the Palomar J-band adaptive optics imaging is plotted as a function of radial distance from the primary target. The filled circles and solid line represent the measured J-band limits; each filled circle represents one step in FWHM. The dashed line represents the derived corresponding limits in the Kepler bandpass based upon the expected Kepmag - J colors (Howell et al. 2011b).Right; As at left, but for the Ks observations.

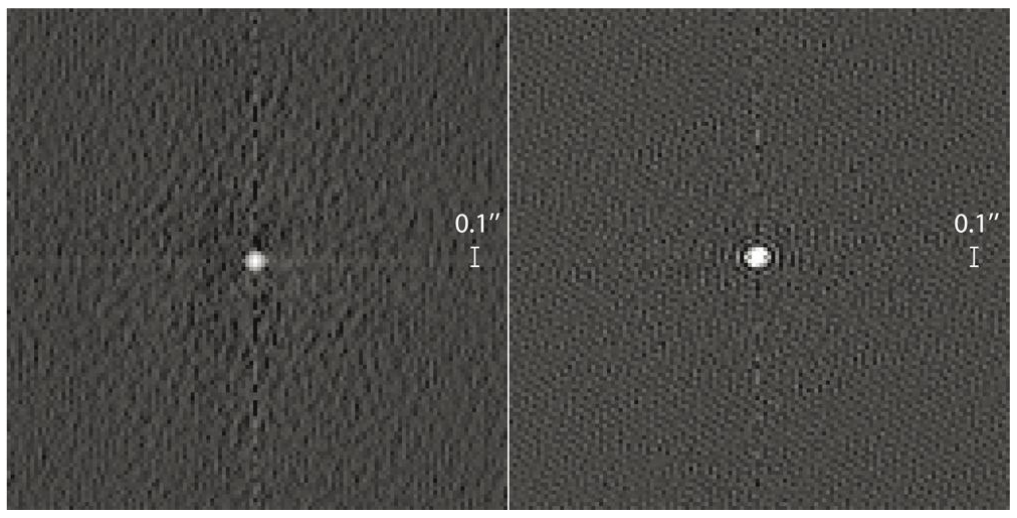

Figure 7. WIYN speckle images. Left, R-band image at 692nm with 40nm passband and Right, I-band at 880nm with 50 nm passband. Field of view is 2.8″ x 2.8″ centered on the star.

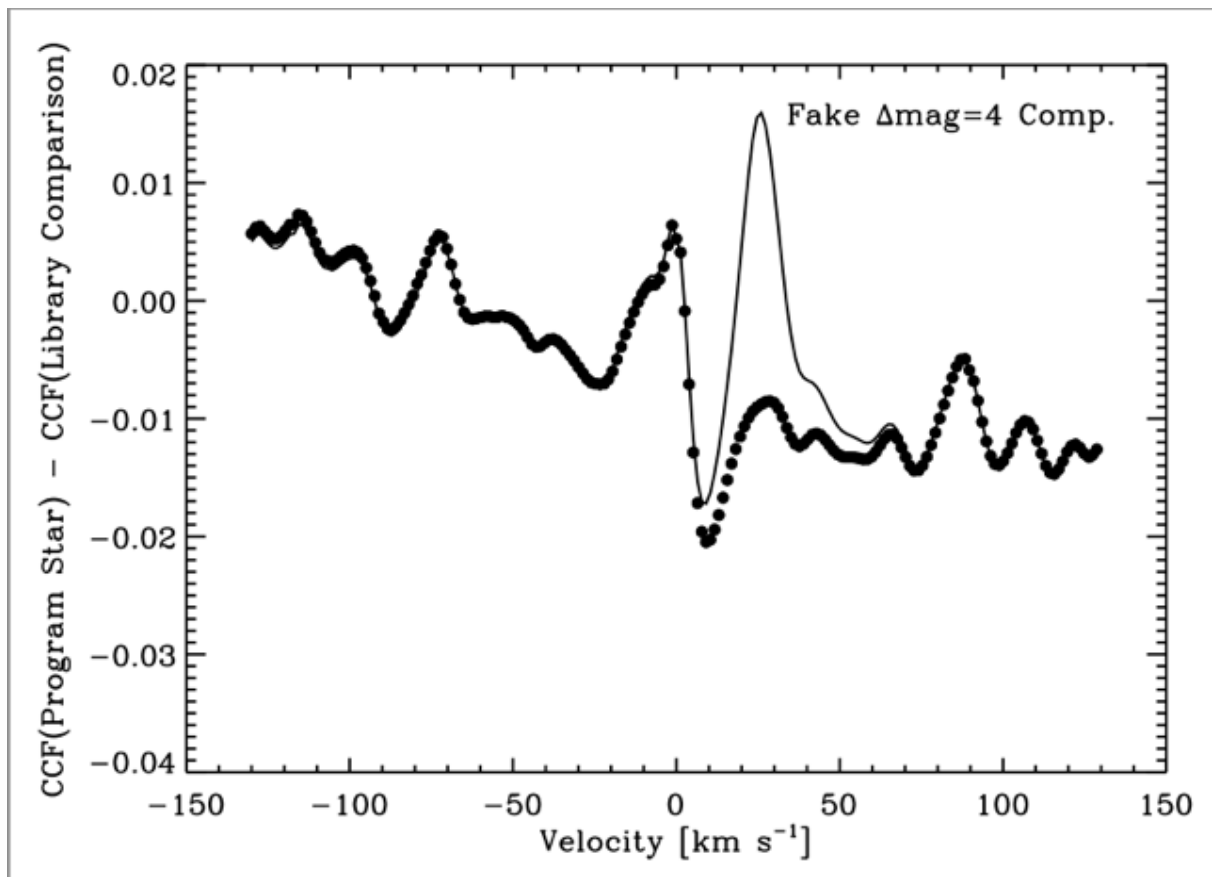

Figure 8. Solid line (without data points); simulation of the effect on the cross-correlation function of adding a "fake" second star 4 magnitudes dimmer than Kepler-22 and with a velocity difference of 30 km $s^{-1}$ to that of the target star. The dots represent the difference in the CCF between that of KOI-87 (against the Solar spectrum) and that of HD 90156 (also against the solar spectrum).

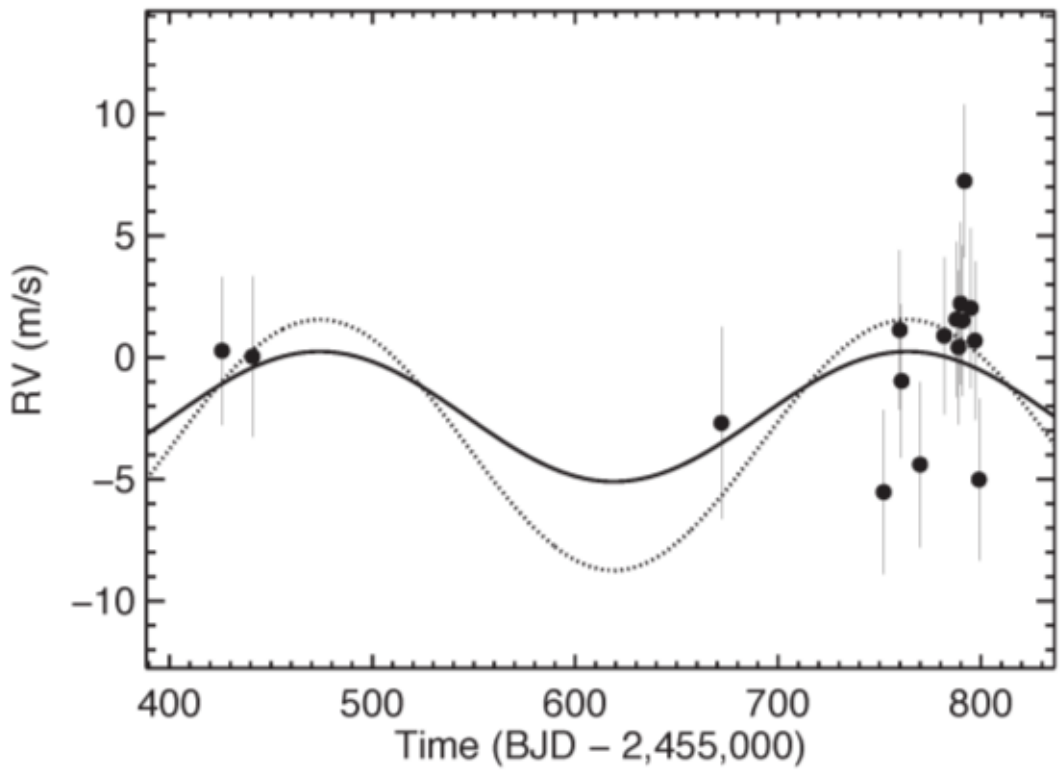

Figure 9: Keck-HIRES RV measurements for Kepler-22 during a year. No significant variation is detected above the typical errors of ~4 m $s^{-1}$, stemming from the internal uncertainties (1.5 m $s^{-1}$) and the stellar noise and instrumental effects (~3 m $s^{-1}$ of jitter). The best-fit circular orbit solution (solid line) has a semi-amplitude of only 1.1 m $s^{-1}$ corresponding to a planet mass of 19 $M_{\oplus}$. An MCMC analysis yields a 1σ upper limit of 27 $M_{\oplus}$ (dashed line) for a planet in the circular orbit based on the photometric-determined epoch and period.

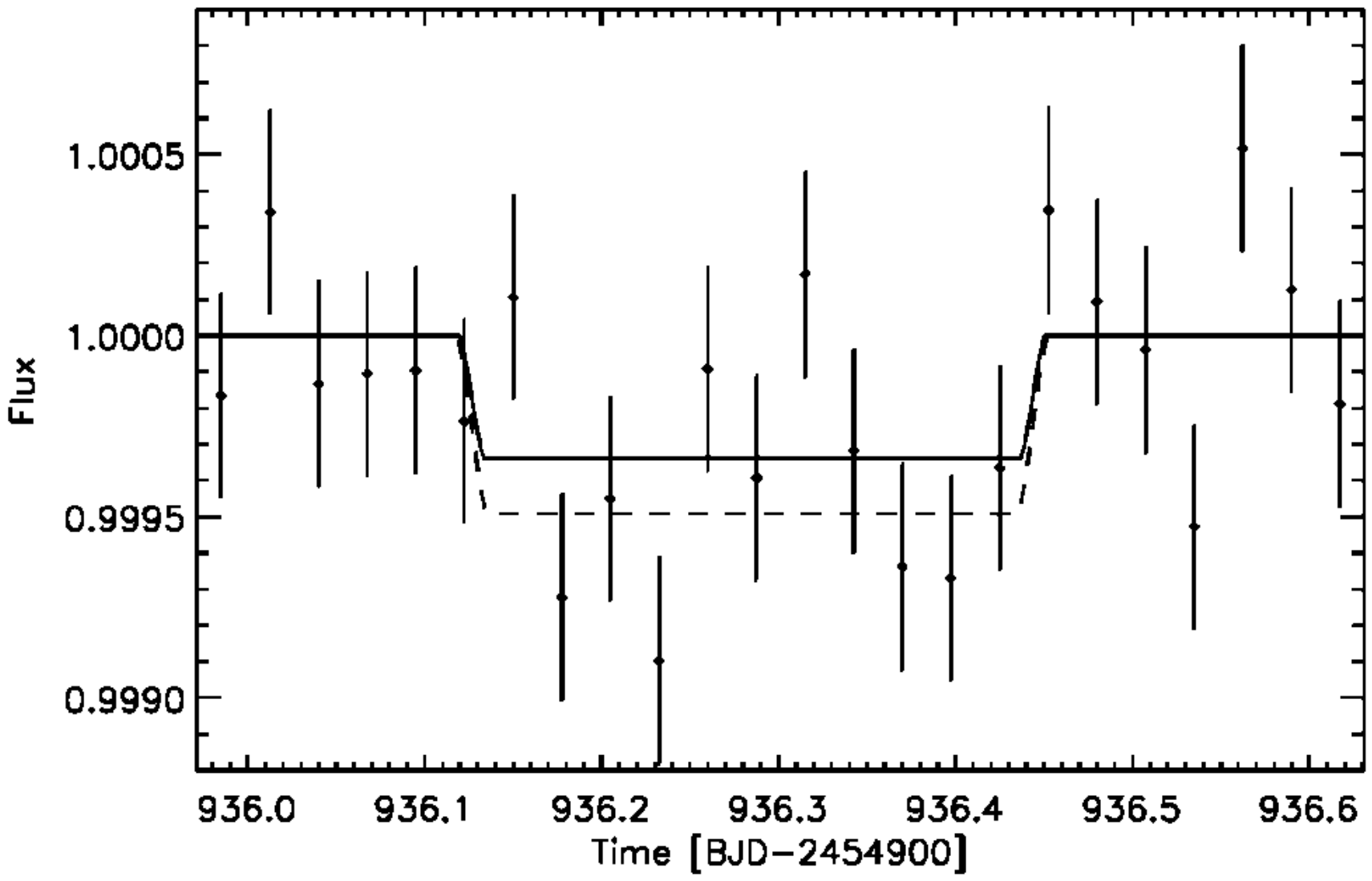

Figure 10. The 1 Oct 2011 transit of Kepler-22b obtained with Warm *Spitzer* at 4.5 μm. The best-fit transit model with depth derived from the Spitzer observations is shown with the solid line, while the *Kepler* transit model (corrected for limb darkening) is shown with a dashed line. The *Spitzer* and *Kepler* transit depths are in agreement (within 1σ).

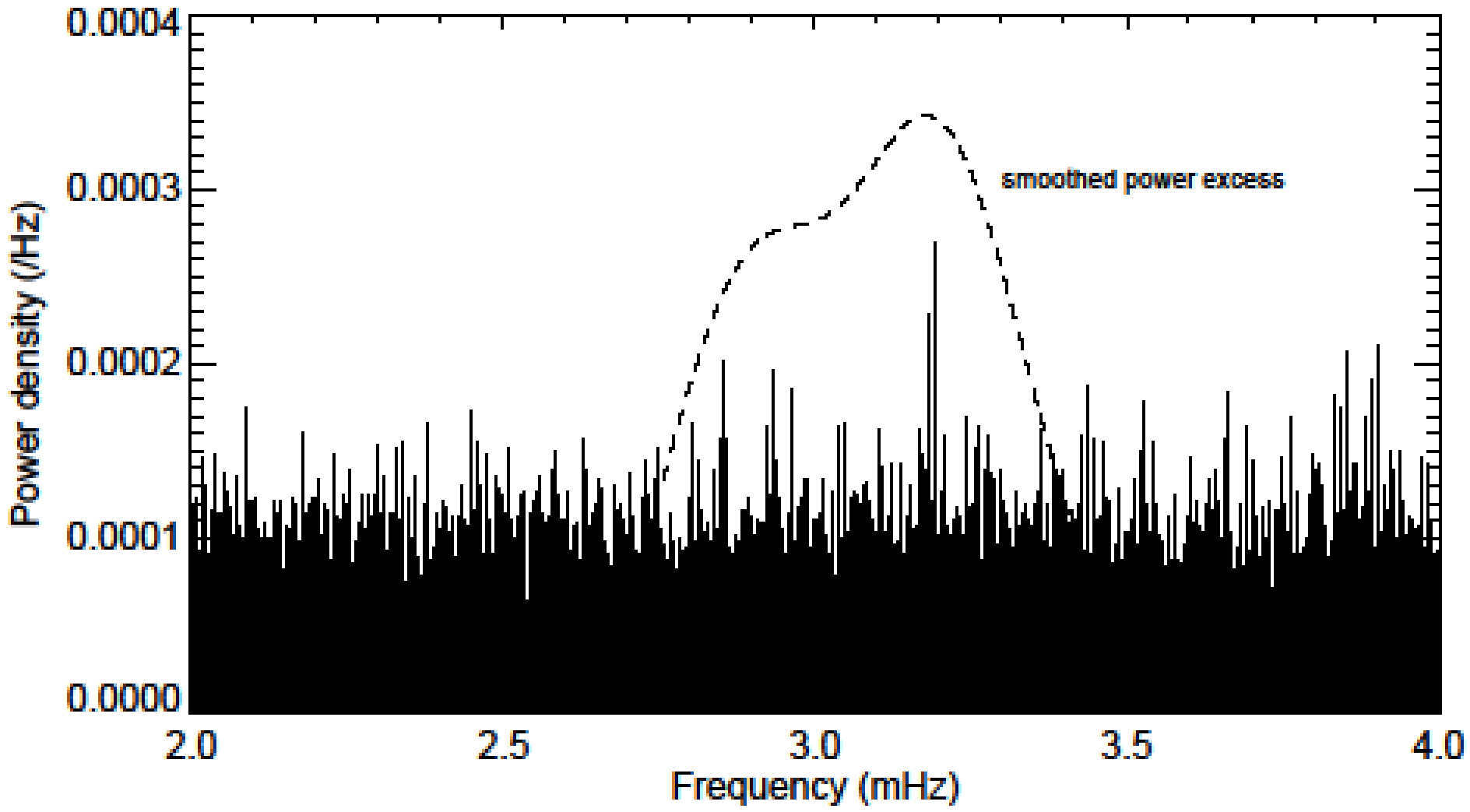

Figure 11. Power spectrum of the times series data for Kepler-22 with 1 minute cadence (19 month of data). Shown is the excess of power near 3 mHz calculated from smoothing the power spectrum (FWHM of filter: 0.4 mHz).

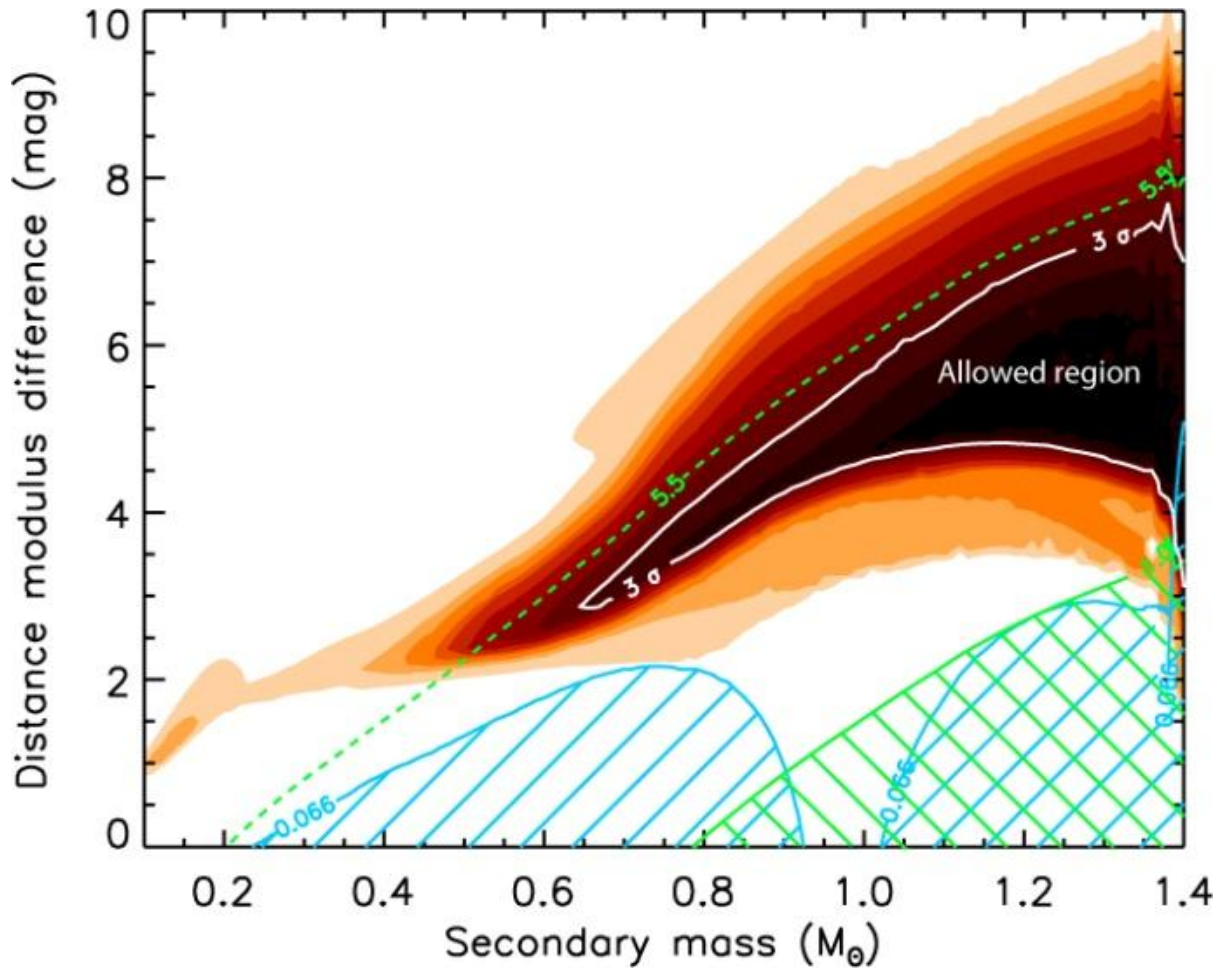

Figure 12. Map of the chi-square surface (goodness of fit) for blends involving background eclipsing binaries composed of two stars. The vertical axis represents the distance between the background pair of objects and the primary star, expressed in terms of the difference in the distance modulus. Only blends inside the solid white contour match the Kepler light curve within acceptable limits (3σ, where σ is the significance level of the chi-square difference compared to a transit model fit; see Fressin et al. 2011). Lighter-colored areas (red, orange, yellow) mark regions of parameter space giving increasingly worse fits to the data (4σ, 5σ, etc.), and correspond to blends we consider to be ruled out. The hatched blue regions at the bottom correspond to blends that can be excluded as well because of their overall r-$K_s$ colors, which are either too red (left) or too blue (right) compared to the measured value for Kepler-22b, by more than 3σ (0.066 mag). The solid diagonal green line is the locus of eclipsing binaries that are 1 mag fainter than the target. Blends in the hatched green area below this line are ruled out because they are bright enough to have been detected spectroscopically. In the case of Kepler-22b, the above color and brightness constraints are redundant with those from BLENDER, which already rules out blends in these areas based on the quality of the light curve fit. Viable blends are all seen to be less than about 5.5 magnitudes fainter than the target (indicated with the dotted green line).

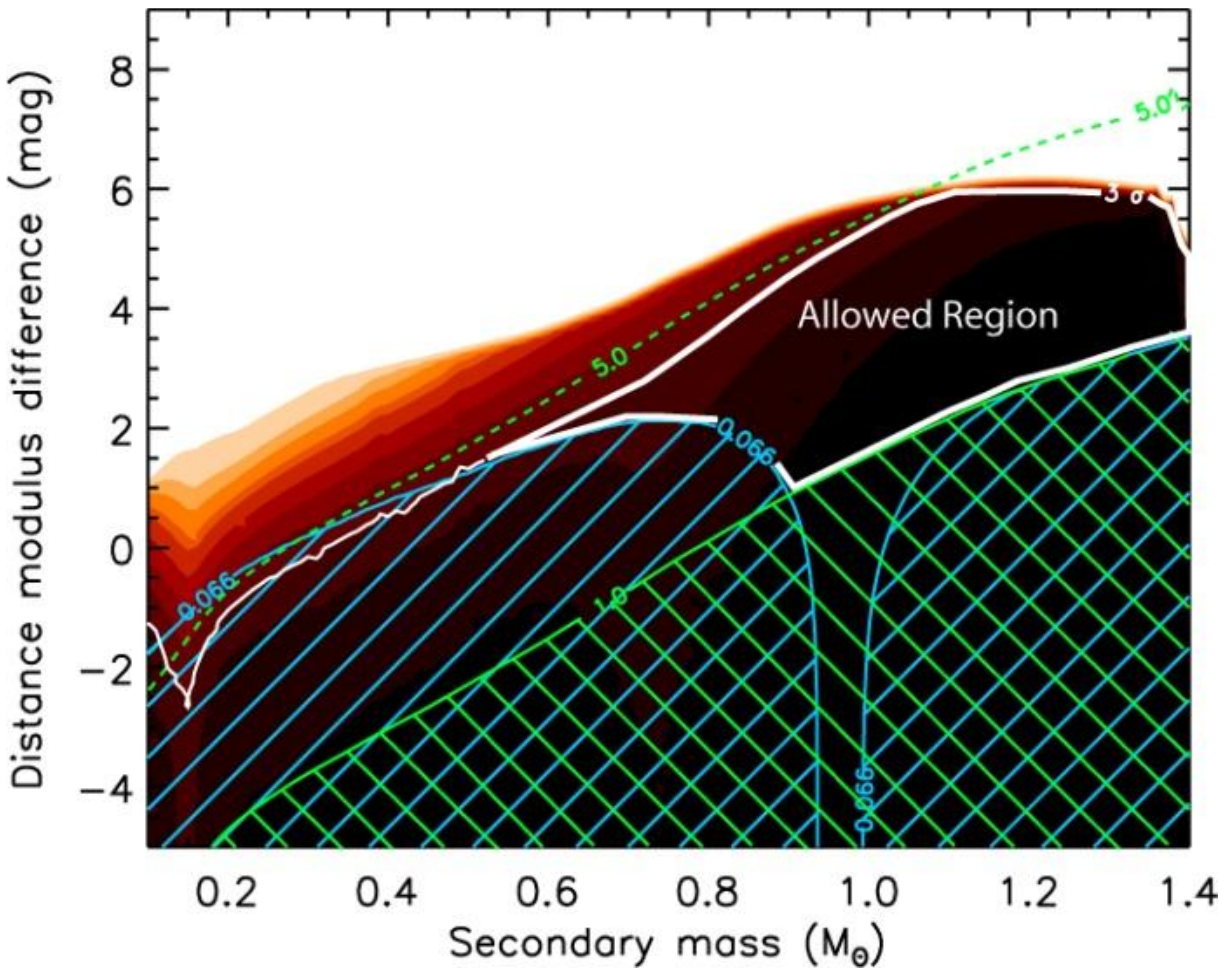

Figure 13. Similar to Figure 12 for blends involving background or foreground stars transited by a larger planet. For this type of blend the color and brightness constraints exclude large portions of parameter space. The only viable blends that remain reside in the area labeled 'Allowed Region', delimited by the thick white contour. These blends are all within about 5 mag of the target (dotted green line).

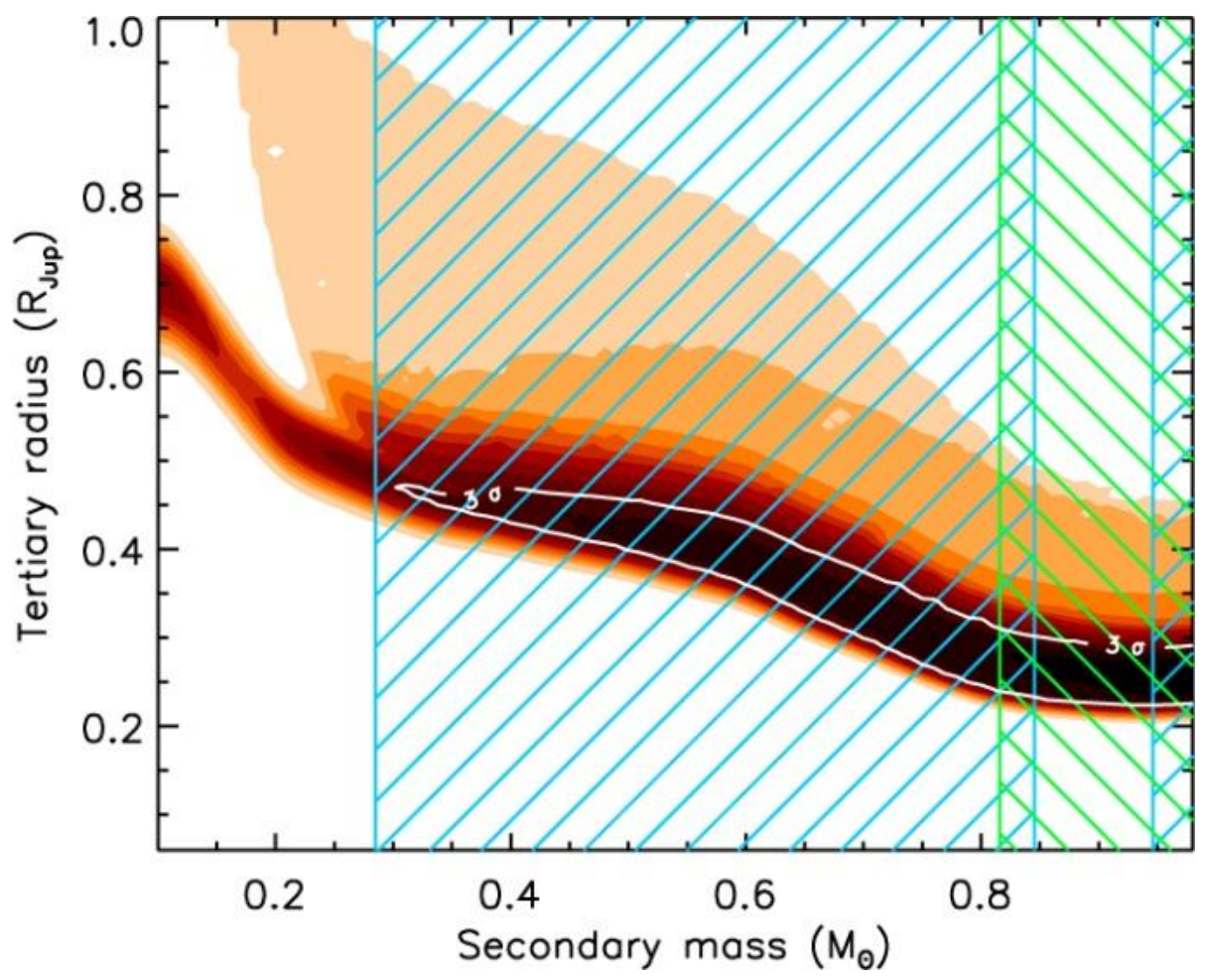

Figure 14**.** Similar to Figures 12 and 13 for the case of hierarchical triple systems in which the secondary is transited by a planet. Blends inside the white 3σ contour yield light curves that match the shape observed for Kepler-22b. However, the combination of the color and brightness constraints (hatched blue and green areas, respectively) exclude all of these false positives.

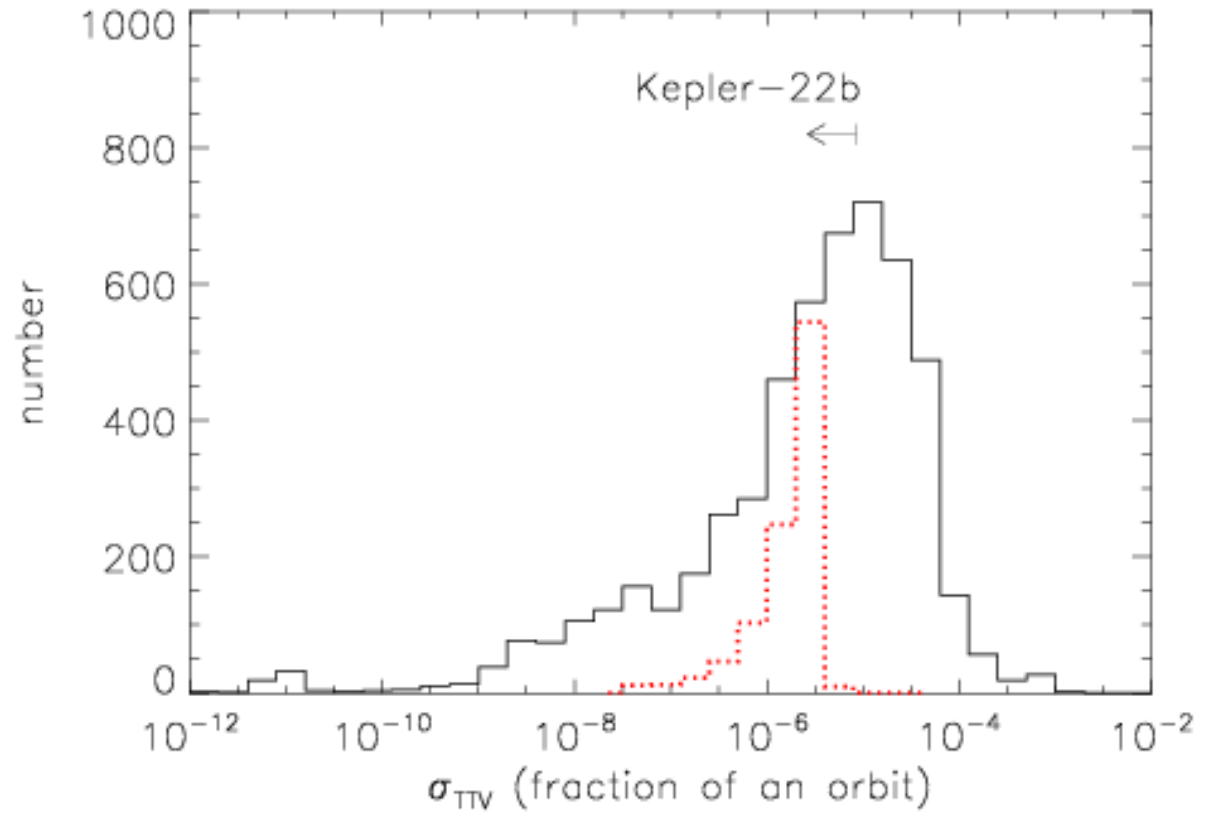

Figure 15. The standard deviation of the timing signal of sets of three adjacent transits. (Solid black:) A histogram for the Super-Earth and Neptune (SEN) planets from Mayor et al. (2011) shows a prominent peak at $10^{-5}$ planetary orbital periods. (Dotted red:) The histogram for Earth itself shows a slightly lower value. (Arrow:) Tail shows the 95% confidence upper limit on the timing RMS of Kepler-22b and tip shows possible position after an 8 year mission. The closeness of this value to known systems demonstrates that Kepler is sensitive to timing variations in the habitable zone.

# Tables

Table 1. Characteristics of Kepler-22 and -22b.

| Parameter | Value |
|---|---|
| Effective temperature, $T_{eff}$ (K) | 5518 ± 44 |
| Surface gravity, log $g$ (cgs) | 4.44 ± 0.06 |
| Metallicity, [Fe/H] | -0.29 ± 0.06 |
| Projected rotation $v$ sin $i$ (km $s^{-1}$) | 0.6 ± 1.0 |
| Density, g $cm^{-3}$ | 1.458 ± 0.030 |
| Mass, $M_{\odot}$ | 0.970 ± 0.060 |
| Radius, $R_{\odot}$ | 0.979 ± 0.020 |
| Luminosity, $L_{\odot}$ | 0.79 ± 0.04 |
| Kepler Magnitude (mag) | 11.664 |
| Age (Gyr) | Not determined |
| Distance (pc) | 190 |
| Orbital period, P (days) | 289.8623 +0.0016/-0.0020 |
| Epoch, T0 (BJD-2454900) | 66.6983 ± 0.0023 |
| Scaled semi-major axis, a/$R_*$ | 186.4+1.1/-1.6 |
| Scaled planet radius, $R_p/R_*$ | 0.0222 + 0.0012/-0.0011 |
| Impact parameter, b (eccentric orbit) | 0.768 + 0.132/-0.078 |
| Orbital inclination, i (degree) | 89.764 + 0.025/-0.042 |
| Transit duration, Δ (hours) | 7.415 + 0.067/-0.078 |
| Radius, $R_{\oplus}$ | 2.38 ± 0.13 |
| Mass, $M_{\oplus}$ ,(1$\sigma$, 2 $\sigma$, & 3$\sigma$ upper limits) | 36, 82, 124 |
| Orbital semi-major axis, $a$ (AU) | 0.849 + 0.018/-0.017 |
| Equilibrium temperature, $T_{eq}$ (K) | 262 |

NOTE. - Uncertainties are standard deviation or +1$\sigma$/-1$\sigma$ unless otherwise noted.

Table 2. Offset of the transit signal source from Kepler-22

| | PRF Fit to Difference Image | Flux-weighted Centroid Motion |
|---|---|---|
| Offset in RA (arcsec) | -0.0525 ± 0.2365 | 0.1080 ± 0.3239 |
| Offset in Dec (arcsec) | 0.2745 ± 0.1916 | -0.1268 ± 0.2980 |
| Offset Distance (arcsec) | 0.2795 ± 0.1934 | 0.1666 ± 0.3091 |
| Offset Distance/$\sigma$ | 1.45 | 0.539 |

Table 3. PRF fit Centroids of Differences and Fit Images

| | Row | Column (pixels) |
|---|---|---|
| Direct Image (pixels) | 615.83 ± 6.03x10$^{-6}$ | 179.86 ± 6.85x10$^{-6}$ |
| Difference Image (pixels) | 615.76 ± 2.11x10$^{-2}$ | 180.01 ± 3.33x10$^{-2}$ |
| Offset (pixels) | -7.03x10-2 ± 2.11x10$^{-2}$ | 1.52e-1 ± 3.33x10$^{-2}$ |
| Offset/$\sigma$ | -3.33 | 4.57 |
| Offset Distance (arcsec) | 1.67e-1 ± 3.15x10$^{-2}$ | |
| Offset Distance/$\sigma$ | 5.32 | |

Table 4. Relative Radial Velocity Measurements of Kepler-22 .

| UT date | BJD-2450000 (days) | RV (m.s$^{-1}$) | Uncertainty (m.s$^{-1}$) | *Chi* | Photons/pixel |
|---|---|---|---|---|---|
| 2010/08/17 | 5425.885914 | 0.27 | 1.23 | 1.17 | 33592 |
| 2010/09/01 | 5440.915924 | 0.04 | 1.57 | 1.11 | 30550 |
| 2011/04/20 | 5672.013942 | -2.69 | 2.29 | 1.078 | 11134 |
| 2011/07/09 | 5751.993081 | -5.53 | 1.67 | 1.126 | 27395 |
| 2011/07/17 | 5759.943140 | 1.13 | 1.55 | 1.163 | 36214 |
| 2011/07/18 | 5760.847325 | -0.96 | 1.37 | 1.15 | 29093 |
| 2011/07/27 | 5769.913318 | -4.39 | 1.69 | 1.119 | 25381 |
| 2011/08/08 | 5781.779644 | 0.89 | 1.48 | 1.108 | 24310 |
| 2011/08/14 | 5787.804765 | 1.57 | 1.43 | 1.132 | 26081 |
| 2011/08/15 | 5788.866202 | 0.42 | 1.4 | 1.118 | 26069 |
| 2011/08/16 | 5789.895579 | 2.21 | 1.62 | 1.084 | 19546 |
| 2011/08/17 | 5790.771600 | 1.51 | 1.29 | 1.121 | 25988 |
| 2011/08/18 | 5791.851531 | 7.24 | 1.35 | 1.143 | 27543 |
| 2011/08/21 | 5794.956506 | 2.02 | 1.55 | 1.104 | 21596 |
| 2011/08/23 | 5796.910381 | 0.69 | 1.52 | 1.119 | 27205 |
| 2011/08/25 | 5798.939125 | -5.01 | 1.6 | 1.088 | 22112 |

NOTE. - *Chi* is the square root of the chi-square statistic describing the sum of the squares of the residuals between the observed spectrum and the model of the spectrum. A low value of "chi" below 2.0 indicates a good fit of the observed spectrum. The radial velocity is one free parameter in that chi-square fit.

Table 5. Estimate of blend frequency for Kepler-22.

| | | Blends involving stellar tertiaries | | | | Blends involving planetary tertiaries | | | |
|---|---|---|---|---|---|---|---|---|---|
| *Kp* range (mag) | $\Delta Kp$ (mag) | Star field density[a] (sq. deg$^{-1}$) | $\rho_{max}$ ('') | Stars ($\times 10^{-6}$) | EBs $f_{EB}$=0.78% ($\times 10^{-6}$) | Star field density (sq.deg$^{-1}$) | $\rho_{max}$ ('') | Stars ($\times 10^{-6}$) | Transiting planets 0.32-2.00$R_{Jup}$, $f_{Plan}$ = 0.28% ($\times 10^{-6}$) |
| 11.7--12.2 | 0.5 | … | … | … | … | … | … | … | … |
| 12.2--12.7 | 1 | … | … | … | … | … | … | … | … |
| 12.7--13.2 | 1.5 | 1 | 0.12 | 0.00349 | 0.0000272 | 86 | 0.12 | 0.3 | 0.00084 |
| 13.2--13.7 | 2 | 15 | 0.12 | 0.0524 | 0.000408 | 158 | 0.12 | 0.552 | 0.00155 |
| 13.7--14.2 | 2.5 | 44 | 0.16 | 0.273 | 0.00213 | 243 | 0.16 | 1.51 | 0.00423 |
| 14.2--14.7 | 3 | 83 | 0.19 | 0.726 | 0.00566 | 386 | 0.19 | 3.38 | 0.00946 |
| 14.7--15.2 | 3.5 | 199 | 0.24 | 2.78 | 0.0217 | 583 | 0.24 | 8.14 | 0.0228 |
| 15.2--15.7 | 4 | 399 | 0.27 | 7.05 | 0.055 | 695 | 0.27 | 12.3 | 0.0344 |
| 15.7--16.2 | 4.5 | 993 | 0.29 | 20.2 | 0.158 | 762 | 0.29 | 15.5 | 0.0434 |
| 16.2--16.7 | 5 | 1402 | 0.31 | 32.7 | 0.255 | 1094 | 0.31 | 25.5 | 0.0714 |
| 16.7--17.2 | 5.5 | 2068 | 0.35 | 61.4 | 0.479 | … | … | … | … |
| 17.2--17.7 | 6 | … | … | … | … | … | … | … | … |
| 17.7--18.2 | 6.5 | … | … | … | … | … | … | … | … |
| 18.2--18.7 | 7 | … | … | … | … | … | … | … | … |
| 18.7--19.2 | 7.5 | … | … | … | … | … | … | … | … |
| 19.2--19.7 | 8 | … | … | … | … | … | … | … | … |
| 19.7--20.2 | 8.5 | … | … | … | … | … | … | … | … |
| Totals | | 5188 | … | 124.46 | **0.977** | 4007 | … | 67.18 | **0.184** |
| Blend frequency (BF) = (0.977 + 0.184) × $10^{-6}$ = 1.16 × $10^{-6}$ | | | | | | | | | |

NOTE. - Magnitude bins with no entries correspond to brightness ranges in which all blends are ruled out by a combination of BLENDER and other constraints.

[a] The number densities in Columns 3 and 7 differ because of the different secondary mass ranges permitted by BLENDER for the two kinds of blend scenarios.

Table 6. Epochs at mid-transit

| Transit Number | Transit time (center of fitted transit) with uncertainties |
|---|---|
| 1 | BJD2454966.69775 +/- 0.00218 |
| 2 | BJD2455256.55988 +/- 0.00185 |
| 3 | BJD2455546.42440 +/- 0.00191 |